\documentclass[a4paper,DIV14]{scrartcl}

\usepackage{a4wide}

\usepackage[utf8]{inputenc}
\usepackage[T1]{fontenc}
\usepackage[english]{babel}
\usepackage{hyperref} 
\usepackage{framed}
\usepackage{color}
\usepackage[usenames,dvipsnames,svgnames]{xcolor}

\usepackage{cite}
\usepackage[ngerman]{varioref}
\usepackage{multirow}
\usepackage{array}
\usepackage[fixlanguage]{babelbib}

\usepackage{graphicx}
\usepackage{amsmath}
\usepackage{amssymb}
\usepackage{amsthm}
\usepackage{stmaryrd}
\usepackage{bbm}

\usepackage[Algorithmus]{algorithm}
\usepackage{algpseudocode}

\vfuzz \hfuzz
\definecolor{LighterGray}{gray}{0.80}

\newcolumntype{x}[1]{>{\centering\hspace{0pt}}p{#1}}

\newcommand{\pointset}[1]{
  \ensuremath{\mathcal{#1}}
}
\newcommand{\setr}{
  \ensuremath{\mathbb{R}}
}
\newcommand{\setn}{
  \ensuremath{\mathbb{N}}
}
\newcommand{\OO}[1]{
  \ensuremath{\mathcal{O}(#1)}
}

\newcommand{\cost}
{
  \ensuremath{\mathrm{cost}}
}

\newcommand{\dist}
{
  \ensuremath{\mathrm{d}}
}
\newcommand{\weight}
{
  \ensuremath{\mathrm{w}}
}
\newcommand{\probdist}
{
  \ensuremath{\mathcal{D}}
}

\newcommand{\ball}
{
  \ensuremath{\mathrm{ball}}
}
\newcommand{\E}
{
  \ensuremath{\mathrm{E}}
}

\newcommand{\etal}{et~al.\ }
\newcommand{\ie}{i.~e.,}
\newcommand{\eg}{e.~g.,}

\newcommand{\method}[1]{
  \text{\textsc{#1}}
}

\algloopdefx{ChooseBest}[1]{\textbf{choose best one out of} #1}

\newcommand{\apln}{5}

\newtheorem{definition}{Definition}
\newtheorem{contract}[definition]{Fact}

\newcommand{\turboprobi}{PROBI}
\newcommand{\plloyd}{P-LLOYD++}

\newenvironment{Bild}
  {\par\raggedbottom\null\vfill\noindent\minipage{\textwidth}\centering}
  {\endminipage\par\vfill\vfill}

\allowdisplaybreaks[1]

\title{PROBI: A Heuristic for the probabilistic $k$-median problem}
\author{Hendrik Fichtenberger and Melanie Schmidt}

\begin{document}

\maketitle


\paragraph*{Abstract.}
We develop the heuristic PROBI for the probabilistic Euclidean $k$-median problem based on a coreset construction by Lammersen \etal~\cite{LSS12}. Our algorithm computes a summary of the data and then uses an adapted version of $k$-means++~\cite{Artkm2007} to compute a good solution on the summary. The summary is maintained in a data stream, so PROBI can be used in a data stream setting on very large data sets.

We experimentally evaluate the quality of the summary and of the computed solution and compare the running time to state of the art data stream clustering algorithms.

\section{Introduction}

Clustering is a basic machine learning task: Partition a set of objects into subsets of similar objects. Geometric clustering is used to cluster sets of points, and the (dis)similarity between the points is then measured by a distance function. Geometric clustering problems then differ by the choice of the distance function. A very natural and popular distance function is the Euclidean $k$-median distance function that measures the dissimilarity between two points by their Euclidean distance, \ie\ the input to this problem consists of points from the Euclidean space $\mathbb{R}^d$. 

The Euclidean $k$-median problem is NP-hard~\cite{MS84}, but it allows for an arbitrarily good approximation with a polynomial time approximation scheme (PTAS). 
The first PTAS was proposed by Arora \etal \cite{ARR98} and used the technique of the famous PTAS for Euclidean TSP by Arora \cite{A98}. 
Subsequently, many PTAS for the Euclidean $k$-median were developed, achieving better and better running times 
\cite{BadHI.App.2002,Chen.On.2009,FeldmanMS.AP.2007,FeldmanL.Au.2011,FrahlingS.Cor.2005,HarPeledK.Sma.2007,HarPeledM.On.2004,KR07,I04,KumarSS.Lin.2010,LangbergS.Uni.2010}. 
The algorithm by Kumar, Sabharwal and Sen \cite{KumarSS.Lin.2010} achieves a running time of $\mathcal{O}(2^{(k/\varepsilon)^{O(1)}} d n)$ which is linear for constant $k$ and $\varepsilon$ and has in particular a notably small dependence on the dimension.

However, in the presence of constantly increasing amounts of data arising from physics, social science or biology, algorithms with linear running time are not always fast enough. Experiments like the Large Hadron Collider generate so much data that even reading them more than once is time consuming. Consequently, \emph{data stream algorithms} have risen to quite some popularity. Here, only one pass over the data is allowed (without any assumptions on the ordering of the data), and the algorithm is only allowed to store a small amount, \eg\ the storage size should be polylogarithmic in the input size or even independent of the length of the stream. 

Among the above cited algorithms, \cite{Chen.On.2009,FeldmanL.Au.2011,FeldmanMS.AP.2007,FrahlingS.Cor.2005,HarPeledM.On.2004,HarPeledK.Sma.2007,I04,LangbergS.Uni.2010} can also be used in a data stream. 
A very popular strategy to develop a streaming algorithm, used by all the cited streaming algorithms except \cite{FrahlingS.Cor.2005} and \cite{I04}, is to design an algorithm which computes a summary of the data and then to turn this algorithm into a streaming algorithm by the use of a technique called Merge \& Reduce. 
This approach leads to an approximation algorithm if the summary is a \emph{coreset}.

In the setting of Euclidean $k$-median clustering, a coreset is a small set $S$ of weighted points from $\mathbb{R}^d$ which meets the following requirement. For every choice of $k$ centers from $\mathbb{R}^d$, the sum of the weighted distances of all points in $S$ to their closest center in $C$ is a $(1+\varepsilon)$-approximation of the sum of the distances of all points in $P$ to their nearest center in $C$. Such a coreset has the nice property that the union of two coresets is a coreset again. 
So a reasonable idea is to partition the input data into chunks, compute a coreset for each chunk and union the resulting coresets.
Whenever this union grows to large, it is reduced by applying the coreset construction again.
However, this might inccur an additional error, so it has to be done a little more careful. This is exactly what the Merge \& Reduce technique does.

Merge \& Reduce goes back to \cite{BenSDec1980} and was first applied to clustering problems in \cite{HarPeledM.On.2004}. By computing coresets of coresets in a tree-like fashion, Merge \& Reduce makes sure that no point takes part in more than $\log n$ reduction steps, thus bounding the error that is incurred. To make up for the (bounded) additional error, Merge \& Reduce requires that the coresets are by a polylogarithmic factor larger than the coresets computed by the original coreset construction.
The currently best streaming algorithm is using this approach~\cite{FeldmanL.Au.2011} and achieves a space complexity of only $\mathcal{O}(k \cdot \log(1/\varepsilon)\cdot \log^4 n / \varepsilon^3)$. 

When implementing algorithms, their theoretical guarantee can sometimes be misleading. This is in particular true for streaming settings and large data settings in general, because high constants in the running time directly make algorithms infeasible for huge amounts of data. Fast heuristics are an alternative, but give no guarantee for good solutions, so the question is whether there exist good trade-offs between practical needs and theoretical guarantees.

This trade-off has been investigated for the related $k$-means clustering problem. For the non streaming version, the $k$-means++ algorithm \cite{Artkm2007} is an improved version of the most famous heuristic for the $k$-means problem, Lloyd's algorithm \cite{L82}, which has a reasonable theoretical guarantee but only adds a short initialization to Lloyd's algorithm and is thus still fast. In the streaming setting, StreamKM++~\cite{AMRSLS12} gives a practical streaming implementation of $k$-means++. It can be proved that Stream-KM++ computes a coreset. On the coreset, it applies $k$-means++. BICO~\cite{FGSSS13} also computes a coreset and uses $k$-means++ on the coreset, but by avoiding the Merge \& Reduce technique and instead building upon a data structure by the famous streaming heuristic BIRCH \cite{ZRL97}, it achieves much better running times than Stream-KM++  while computing solutions of the same quality.

We further investigate the trade-off between worst-case guarantees and empirical performance in the setting of \emph{probabilistic} clustering. In addition to being large, nowadays data often comes with uncertainty. 
For example, data collected by sensors usually has faults. But also data which is originally precise can lead to data sets with uncertainty, for example if data bases are joined and it is not certain which entities in the different data bases are the same.

The problem to cluster uncertain data has recently triggered the development of new heuristics~\cite{CCKN06,GKS10,KP05_a,KP05_b,NKCC+06,XL08}. In particular, \cite{CCKN06} and \cite{NKCC+06} generalize Lloyd's algorithm to a probabilistic setting. For a survey on different heuristic approaches, see the paper by Aggarwal and Yu~\cite{AY09}.
A theoretic study of probabilistic clustering was done by Cormode and McGregor\cite{CormodeM.App.2008}. They investigate different clustering formulations, in particular they give a $(3+\varepsilon)$-approximation algorithm for the probabilistic Euclidean $k$-median problem. Guha and Munagala improve one of their results on so-called $k$-center clustering~\cite{GM09}.

For probabilistic clustering, only one coreset construction is known. Lammersen \etal \cite{LSS12} give a reduction for general metric $k$-median clustering to the deterministic case. Of course this would also work in the Euclidean setting  because the Euclidean distance is a metric, but the reduction would only reduce it to a deterministic metric $k$-median problem, and those usually assume a finite set as the ground set for possible centers. Thus, \cite{LSS12} additionally includes an algorithm specifically for the Euclidean $k$-median problem, which is a generalization of the coreset construction by Chen~\cite{Chen.On.2009}. 

However, their algorithm consists of many nested subroutines provided by previous results from clustering theory, and implementing it does not immediately result in an efficient or even reasonably fast algorithm.
This report deals with the question how the algorithm in \cite{LSS12} can be heuristically modified in order to be implementable. As there is no theoretical justification for the modifications, we do not expect optimal results. However, we believe that our implementation is a good step in the development of an efficient algorithm for the probabilistic Euclidean $k$-median problem.

\section{Preliminaries}

First, we define the \emph{weighted probabilistic Euclidean k--median--clustering problem}. Notice that Cormode and McGregor \cite{CormodeM.App.2008} refer to our scenario as the \emph{assigned} Euclidean $k$-median problem. 
Let $\pointset{X} := \{ x_1,\ldots,x_m \} \subset \setr^d$ be a finite set of $m$ points from the d--dimensional Euclidean space $\setr^d$, and let $\pointset{V} := \{ v_1,\ldots,v_n \}$ be a set of $n$ \emph{nodes}, where each node $v_i$ follows an independent probability distribution $\probdist_i$ over $\pointset{X}$.
For any $i \in [n]$ and any $j \in [m]$, we denote the probability that the node $v_i$ is realized at $x_j$ by $p_{ij}$.
We denote the total probability that $v_i$ is realized by $p_i := \sum^m_{j=1} p_{ij}$.
We assume that $p_i \leq 1$, which means that with probability $1-p_i$ the node $v_i$ is not realized.

\begin{definition}[Weighted probabilistic Euclidean k--Median~\cite{CormodeM.App.2008}]
  Given $k \in \setn$ and a positive weight function $\weight : \pointset{V} \rightarrow \setr_{\geq 0}$ on the set of nodes $\pointset{V}$, the \emph{weighted probabilistic Euclidean k--median--clustering problem} for the set of nodes $\pointset{V}$ is to find a set \mbox{$\pointset{C} := \{ c_1,\ldots,c_k \} \subset \setr^d$ of $k$} cluster centers and an assignment $\rho : \pointset{V} \rightarrow \pointset{C}$ such that the expected k--median clustering cost
  \begin{equation*}
    \E[\cost_{\weight}(\pointset{V},\pointset{C},\rho)] := \sum_{i=1}^n w_i \sum_{j=1}^m p_{ij} \, \dist(x_j, \rho(v_i))
  \end{equation*}
  is minimized.
\end{definition}

For a given instance, let $p_{min}$ be the smallest realization probability, i.e. $\min_{v_i \in \pointset{V}, x_j \in \pointset{X}} p_{ij}$, and let $w_{min}$ be the minimum of the smallest weight and $1$, i.e. $w_{min} := \min \{ \min_{v_i \in \pointset{V}} w(v_i), 1\}$.
We denote the expected weight of all nodes by $W := \sum_{v_i \in \pointset{V}}{\weight(v_i)p_i}$.

Next, we give a definition of a \emph{coreset} for the weighted probabilistic Euclidean k--median--clustering problem.
Our definition restricts both the number of nodes in the coreset and the size of the probability distributions describing these points.
Let $\pointset{U} := \{ u_1,\ldots,u_s \}$ be a set of $s$ nodes where each $u_o \in \pointset{U}$ follows an independent probability distribution $\probdist'_o$ over $\pointset{X}$.
For any $o \in [s]$ and any $j \in [m]$, we denote the probability that $u_o$ is realized at $x_j$ by $p'_{oj}$.
We denote the total probability that $u_o$ is realized by $p'_o := \sum_{j=1}^m p'_{oj}$.

\begin{definition}[Coreset for weighted probabilistic Euclidean k--Median~\cite{LSS12}]
  Given the set of nodes $\pointset{V}$, let $\pointset{U} : \{ u_1,\ldots,u_s \} \subseteq \pointset{V}$ be a weighted set of nodes with positive weight function $\weight' : \pointset{U} \rightarrow \setr_{\geq 0}$, and let $\probdist' := \{ \probdist'_1,\ldots,\probdist'_s \}$ be a set of $s$ probability distributions over $\pointset{X}$ defining the distribution of nodes in $\pointset{U}$.
  Given $k \in \setn$, a positive weight function $\weight : \pointset{V} \rightarrow \setr_{\geq 0}$ on the set of nodes $\pointset{V}$ and a precision parameter $\epsilon$, $0 < \epsilon \leq 1$, the set $\pointset{U}$ is called $(k,\epsilon)$--coreset of $\pointset{V}$ for the weighted probabilistic Euclidean k--median--clustering problem if, for each $\pointset{C} \subset \setr^d$ of size $|\pointset{C}| = k$, we have
  \begin{equation*}
    \left| \min_{\rho : \pointset{U} \rightarrow \pointset{C}} \E_{\probdist'} [ \cost_{w'}(\pointset{U},\pointset{C},\rho)] -
      \min_{\rho : \pointset{V} \rightarrow \pointset{C}} \E_{\probdist'} [ \cost_{w}(\pointset{V},\pointset{C},\rho)] \right| \leq
    \epsilon \cdot \min_{\rho : \pointset{V} \rightarrow \pointset{C}} \E_{\probdist'} [ \cost_{w}(\pointset{V},\pointset{C},\rho)].
  \end{equation*}
\end{definition}

Bicriteria approximations are a widely used relaxation of typical approximation algorithms, where in addition to the quality, the number of centers $k$ does not have to be matched exactly but only approximately.
We give a formal definition of a bicriteria approximation for the weighted probabilistic k--median--clustering problem.
\begin{definition}[Bicriteria approximation]
  Given $k \in \setn$, a positive weight function $\weight : \pointset{V} \rightarrow \setr_{\geq 0}$ and $\alpha, \beta \geq 1$, $\pointset{A}$ is referred to as $[\alpha,\beta]$--approximation of the optimal center set $\pointset{C}$, if
  \begin{align*}
    \min_{\rho : \pointset{V} \rightarrow \pointset{C}} \E_{\probdist'} [ \cost_{w}(\pointset{V},\pointset{A},\rho)] &\leq \alpha \min_{\rho : \pointset{V} \rightarrow \pointset{C}} \E_{\probdist'} [ \cost_{w}(\pointset{V},\pointset{C},\rho)]\\
    |\pointset{A}| &\leq \beta k.
  \end{align*}
\end{definition}

This report is organized as follows. In Section~\ref{sec:originalalgo}, we describe the coreset construction for the Euclidean probabilistic $k$-median problem by Lammersen, Schmidt and Sohler~\cite{LSS12}. In Section~\ref{sec:Implementierung}, we describe how this algorithm can be modified in order to be efficiently implementable. In Section~\ref{sec:Experimente}, we evaluate the implementation empirically.

\section{Coreset for probabilistic Euclidean k--Median}
\label{sec:originalalgo}

In this section, we review the algorithm by Lammersen, Schmidt and Sohler~\cite{LSS12} to compute a coreset for the probabilistic Euclidean $k$-median problem in data streams.
First, we describe the actual coreset construction. Second, we recall how to embed the coreset construction into a data stream setting in order to compute a coreset in a stream.

\subsection{Coreset construction}
\label{sec:ImplKernmengenberechnung}

The coreset construction in~\cite{LSS12} consists of five steps.
We give a detailed pseudo code as Algorithm~\ref{alg:Coreset}.
\begin{enumerate}
	\item Construct a set $\pointset{Y}$ containing a 2--approximation of the probabilistic 1--Median for each node $v_i \in \pointset{V}$.
  \item Construct a set $\pointset{Y} \subseteq \pointset{A}$ which is the center set of an $[\alpha,\beta]$--bicriteria approximation for the metric k--Median problem of $\pointset{Y}$.
  \item Partition $\pointset{Y}$ into buckets $\pointset{Y}_{\ell,h,a}$ that group 1--Medians (i.e. nodes) which are similar in terms of their location and their contribution to the total clustering cost.
  \item Draw a sample $\pointset{U}_{\ell,h,a} \subseteq \pointset{Y}_{\ell,h,a}$ uniformly at random and with replacement from each partition $\pointset{Y}_{\ell,h,a}$.
  \item Approximate the probabilty distribution of each node in $\pointset{U} := \bigcup \pointset{U}_{\ell,h,a}$ by computing a coreset of each node $v_i \in \pointset{U}$
\end{enumerate}

Following these steps, one obtains a $(1+\epsilon)$--coreset of the input $\pointset{V}$ with error probability~$\delta$.
The following algorithms are used in \cite{LSS12} to implement steps 1, 2 and 5:
\begin{description}
  \item[Step 1] The algorithm described in \cite{KumarSS.Lin.2010} by Kumar \etal is used to obtain a 2--approximation of the probabilistic 1--Median for a node $v_i$.
  It runs in constant time, i.e. the number of points $x_j$ with $p_{ij} > 0$ does not affect the running time when processing node $v_i$.
  We denote this algorithm by \method{KumarMedian::approximateOneMedian}.
  \item[Step 2] Two algorithms are used.
  An $[\alpha,\beta]$--bicriteria approximation algorithm for the metric k--Median problem is sped up by an algorithm proposed by Indyk \cite{PiotrIndyk.Sub.1999}.
  This algorithm is denoted by \method{Indyk::computeCenterSet}.
  \item[Step 5] The reduced probability distribution for a node $v_i$ is computed by using an algorithm by Chen \cite{Chen.On.2009}.
  We refer to this algorithm as \method{Chen::computeCoreset}.
\end{description}

\begin{contract}[see \cite{LSS12}]
  Algorithm \ref{alg:Coreset} can be implemented to run in
  \begin{equation*}
    \label{eq:runtime}
    \mathcal{O} \left(knm \log \left(\frac{n}{\delta} \right) \log \left(\log \left(\frac{W}{w_{min} \, p_{min} \, \epsilon} \right) \right) \right)
  \end{equation*}
  time.
\end{contract}

\begin{algorithm}
  \caption{Coreset}
  \label{alg:Coreset}
  \begin{algorithmic}[\apln]
    \Function{computeCoreSet}{$\pointset{X},\pointset{V},\weight,k,\epsilon$}
      \ForAll{$v_i \in \pointset{V}$} \Comment Construct $\pointset{Y}$
        \State $v'_i \gets \emptyset$ \label{alg:cs_gewichtung0}
        \ForAll{$x_j : p_{ij} > 0$}
          \State $v'_i \gets v'_i \cup \{x_j \, | \, n \in \left[ \, \lfloor w(v_i) p_{ij} / (w_{min} p_{min} \epsilon) \rfloor \, \right] \}$
        \EndFor \label{alg:cs_gewichtung1}
        \ChooseBest{$\overline{r}_1$}
          \State $y_i \gets \method{KumarMedian::approximateOneMedian}(v'_i, 0.9)$
        \State $\weight(y_i) \gets \weight(v_i)$
      \EndFor
      \State $\pointset{Y} \gets \bigcup y_i$
      
      \ForAll{$y_i$} \Comment Construct $\pointset{A}$
        \State $y'_i \gets \{y_i \, | \, n \in \left[ \, \lfloor w(v_i) p_i / (w_{min} p_{min} \epsilon) \rfloor \, \right] \}$
      \EndFor
      \State $\pointset{Y}' \gets \bigcup y'_i$
      \ChooseBest{$\overline{r}_2$}
        \State $\pointset{A} \gets \method{Indyk::computeCenterSet}(\pointset{Y}',k)$      
        
      \For{$\ell \in \{1,\ldots,\tau\}$} \Comment Partition \pointset{Y}
        \State $\pointset{Y}_\ell \gets \{y \in \pointset{Y} \, | \arg\min_{i \in [\tau]} \dist(y,a_i) = a_\ell \}$
        \For{$h \in \{0,\ldots,\nu\}$}
          \State $\pointset{Y}_{\ell,h} \gets
                    \begin{cases}
                      \pointset{Y}_{\ell} \cap \ball(a_\ell, R) & h=0\\
                      \pointset{Y}_{\ell} \cap [\ball(a_\ell,2^h R) \, \backslash \, \ball(a_\ell,2^{h-1}R)] & h \geq 1
                    \end{cases}$
          \State \Comment $ball(p,r)$ is the open ball of radius $r$ centered at $p$
          \For{$a \in \{0,\ldots,\nu\}$}
            \State $\pointset{Y}_{\ell,h,a} \gets
                      \begin{cases}
                        \{ y_i \in \pointset{Y}_{\ell,h} \, | \, \sum_{x_j \in \pointset{X}}{(p_{ij}/p_i) \, \dist(x_j,y_i)} \leq R \} & a=0\\
                        \{ y_i \in \pointset{Y}_{\ell,h} \, | \, 2^{a-1} < \sum_{x_j \in \pointset{X}}{(p_{ij}/p_i) \, \dist(x_j,y_i)} \leq 2^a R \} & a \geq 1
                      \end{cases}$
            
          \EndFor
        \EndFor
      \EndFor
      \State $\pointset{U}_{\ell,h,a} \gets \method{weightedSamplingWithReplacement}(\pointset{V}_{\ell,h,a}, s)$
      \State $\pointset{U} \gets \bigcup_{\ell,h,a}{\pointset{U}_{\ell,h,a}}$
      
      \ForAll{$u_j \in \pointset{U}$} \Comment Approximate probability distributions
        \State $u'_j \gets \text{Unweighted multiset: see lines \ref{alg:cs_gewichtung0}--\ref{alg:cs_gewichtung1}}$
        \State $\hat{u}_j \gets \method{Chen::computeCoreset}(u'_j, 1, \epsilon, \delta \, / \, n)$
      \EndFor
      \State $\pointset{\hat{U}} \gets \bigcup_j \hat{u}_j$
      
      \State \Return $\pointset{\hat{U}}$
    \EndFunction
  \end{algorithmic}
\end{algorithm}

\subsection{Streaming Algorithm}\label{sec:streaming}

Lammersen \etal\ use the Merge \& Reduce technique \cite{BenSDec1980,HarPeledM.On.2004} to enable Algorithm~\ref{alg:TurboCoreset} to work in a data stream.
Here, $\pointset{V}$ is given as a stream of $n$ weighted nodes.
Each node $v_i$ is given as a consecutive chunk in the data stream that is a sequence of up to $m$ point--probability pairs in worst case order representing the discrete probability distribution $\probdist_i$ of the node $v_i$.
More precisely, the nodes are organized in a small number of coresets, each representing $2^\ell N$ nodes (for some integer $\ell$ and a fixed constant $N$).
Every time when two coresets representing the same number of nodes exist, we take the union (merge) and create a new coreset (reduce).
The construction is based on the following fact:

\begin{contract}
  \begin{enumerate}
    \item If $\pointset{U}_1$ and $\pointset{U}_2$ are $(k,\epsilon)$--coresets for disjoint sets $\pointset{V}_1$ and $\pointset{V}_2$, then $\pointset{U}_1 \cup \pointset{U}_2$ is a $(k,\epsilon)$--coreset for $\pointset{V}_1 \cup \pointset{V}_2$.
    \item If $\pointset{U}_1$ is a $(k,\epsilon_1)$--coreset for $\pointset{U}_2$ and $\pointset{U}_2$ is a $(k,\epsilon_2$--coreset for $\pointset{U}_3$, then $\pointset{U}_1$ is a $(k,(1+\epsilon_1)(1+\epsilon_2)-1)$--coreset for $\pointset{U}_3$.
  \end{enumerate}
\end{contract}

The idea is as follows.
We maintain buckets $B_0,B_1,\ldots$ which are created on demand.
Bucket $B_0$ can store between $0$ and $N$ nodes.
For $\ell \geq 1$, bucket $B_\ell$ is either empty or stores a coreset $U_\ell$ of approximately $N$ coreset nodes representing $2^{\ell-1}$ nodes from the data stream.
Next, we explain the method in detail.

All nodes in the data stream are processed in the same way.
Let $v_i$ be the $i$--th node read from the data stream.
Then, $v_i$ is inserted into bucket $B_0$.
If bucket $B_0$ is full, then all nodes from $B_0$ are moved to bucket $B_1$.
If bucket $B_1$ is empty, we are done.
Otherwise, we compute a coreset $\pointset{U}_2$ from the union of the approximately $2N$ nodes stored in $B_0$ and $B_1$ using algorithm \ref{alg:TurboCoreset}.
Afterwards, both buckets $B_0$ and $B_1$ are emptied and the approximately $N$ coreset nodes from $\pointset{U}_2$ are moved into bucket $B_2$.
If $B_2$ is empty, we are done.
Otherwise, we compute a coreset $U_3$ from the union of the approximately $2N$ coreset nodes stored in $B_1$ and $B_2$ using algorithm \ref{alg:TurboCoreset}.
Then, buckets $B_1$ and $B_2$ are emptied and the approximately $N$ coreset nodes from $\pointset{U}_3$ are moved into bucket $B_3$.
If bucket $B_3$ is empty, we are done.
Otherwise, we repeat this process until we reach an empty bucket.
The final coreset is constructed by reducing $B_0$ and returning the union of all buckets.

\section{Implementation and Runtime Improvements}
\label{sec:Implementierung}

Now, we describe how we modify Algorithm~\ref{alg:Coreset} with the goal of a practically efficient algorithm. The implementation will be available at http://cgl.uni-jena.de/Software/WebHome.

As this is the first data stream algorithm for this problem, we do not expect a perfectly behaving algorithm. 
However, we would like to achieve that the algorithm is efficient at least for inputs with sparse or only moderatly dense nodes. 
This in particular means that we concentrate on the core algorithm and ignore step 5 for now.

Algorithm~\ref{alg:Coreset} contains several subproblems that have to be solved, and \cite{LSS12} points to asymptotically efficient algorithms for all of them.
However, these algorithms are not necessarily efficient in practice. 
For example, the approximation algorithm to compute the 1-median has an asymptotic running time of $\OO{1}$, but the constant is quite large.
Especially for sparse nodes, introducing a very large constant for every 1-median computation heavily restricts the number of nodes that can be processed in a given amount of time.
The situation is similar for the bicriteria approximation algorithm.

Notice that because the coreset construction is used within a Merge \& Reduce framework, we know an upper bound on the input size for each subproblem.
We will restrict the size of the chunks even more 
and use subroutines that work well on inputs of this known size.

In the following, we describe heuristic approaches to replace the steps of Algorithm~\ref{alg:Coreset} in order to enable it to work for inputs of moderate up to huge size.

\subsection{Construction of $\pointset{Y}$}
\label{sec:ConstructSetY}

Finding the $1$-median or \emph{geometric median} of a point set is also known as the \emph{Fermat-Weber problem} and has a long history. 
In fact, it is impossible to construct the $1$-median using only a ruler and a compass (\emph{straight-edge and compass constructions})~\cite{M73}, and  the problem is not solvable by \emph{radicals}~\cite{B88}, \ie\ it is not expressable in terms of $(+,-,\ast,/,\sqrt[k]{\ \cdot \ })$ over $\mathbb{Q}$. 

One popular approach to approximately determine the 1-median is an iterative approach known as \emph{Weiszfeld's algorithm}\cite{WeiSur,WeiOn2009}. 
It defines a sequence $y^{(k)}$ which converges to the 1--Median of the point set $\pointset{X}$ \cite{KulAN1962} and is defined by 

\begin{equation*} \left.
  y^{({k+1})} =
  \left ( \sum_{x_j \in \pointset{X}} \frac{1}{\Vert x_j - y^{(k)} \Vert} \; x_j \right ) \right /
  \left ( \sum_{x_j \in \pointset{X}} \frac{1}{\Vert x_j - y^{(k)} \Vert} \right ).
\end{equation*}

In the context of our problem to compute the set $\pointset{Y} = \{y_1,\ldots,y_n\}$ of 1--Medians of all nodes $v_i$, we define the recurrence $y_i^{(\nu)}$ for every node $v_i$ by

\begin{equation} \left.
  y^{({\nu+1})}_i =
  \left ( \sum_{x_j \in \pointset{X}} \frac{w_{ij}}{\Vert x_j - y^{(\nu)}_i \Vert} \; x_j \right ) \right /
  \left ( \sum_{x_j \in \pointset{X}} \frac{w_{ij}}{\Vert x_j - y^{(\nu)}_i \Vert} \right ).
  \label{eq:Weiszfeld}
\end{equation}

For each node $v_i$, its 1--Median $y_i$ is approximated as follows:
We choose the center of gravity $(\sum_{x_j \in \pointset{X}} w_{ij} x_j) \, / \, (\sum_{x_j \in \pointset{X}} w_{ij})$ as initial point $y^0_i$.
As long as 
\begin{equation}
  \Vert y^{({\nu})}_i - y^{({\nu-1})}_i \Vert \; / \; \Vert y^{({\nu-1})}_i - y^{({\nu-2})}_i \Vert \leq 0.1 \text{ and } k<15
\label{eq:WeiszfeldStop}
\end{equation}
is satisfied, the successor $y^{\nu+1}_i$ of $y^\nu_i$ is computed by evaluating (\ref{eq:Weiszfeld}) and $k$ is incremented by one.
Finally, when (\ref{eq:WeiszfeldStop}) is violated, we conclude by returning $y^\nu_i$.
The number of iteration $k$ is chosen experimentally. In fact, on the present data the process usually computes good solutions after very few iterations.

There is one special case we have to take care of: The iteration may arrive at a point $y^{(\nu)}_i = x_j \in \pointset{X}$, thus no successor is defined (because $\Vert x_j - y^{(\nu)}_i \Vert$ equals zero).
Since this happens very rarely, we abort the iteration and use Kumar's algorithm as fallback (see section \ref{sec:ConstructSetY}) in this case.
Notice that for non-sparse nodes, using Kumar's algorithm is a good idea in any case because of its (high but) constant running time. 

We will refer to this algorithm as $\method{Weiszfeld::approximateOneMedian}$.

\subsection{Construction of \pointset{A}}
\label{sec:LloydMedian}

Lloyd's algorithm \cite{L82} for the $k$-means problem is probably the most used clustering algorithm.
Algorithms of similar structure are also used for other clustering functions, and we adapt it to the probabilistic Euclidean $k$-median problem.

The original algorithm starts with an initial set of $k$ centers from $\mathbb{R}^d$ and then alternately performs two steps until a good enough solution is found.
Step 1 assignes every point to its closest center in the current center set $S$. Step 2 computes the center of gravity for all subsets and replaces $S$ by the set of these $k$ centers.
Notice that both steps can only improve the cost, because assigning each point to its closest center can only be cheaper than assigning it to any other center, and for given subsets of points with the same center, the center of gravity is the optimal choice as a center.
However, there are inputs where the $k$-means algorithm needs an exponential number of iterations, and it is also known that it may converge to a local optimum.

The $k$-means++ \cite{Artkm2007} algorithm improves this behaviour by adding a procedure to compute a good initial set of centers. It chooses an initial set iteratively, starting with one point chosen uniformly at random. 
In every step, the distance of each point to the so-far chosen points is computed. Then, a point is chosen randomly and added to the center set, and the probability of each point to be chosen is proportional to the squared distance of the point.

To adapt this algorithmic idea for the (deterministic) Euclidean $k$-median problem, we need a subroutine to compute a 1-median of a subset of points. As in Section~\ref{sec:ConstructSetY}, we use Weiszfeld's algorithm to find an estimate for the 1-median. This might not result in the best 1-median, but should on average produce a 1-median that is better than the previous center of the subset. 

The adaptation of the $k$-means++ seeding procedure is straightforward, we choose $k$ centers iteratively and base the probability distribution on the Euclidean $k$-median cost instead of the squared Euclidean distance. 
We get the following algorithm, which we call $\method{LloydMedian::computeCenterSet}$:

\begin{enumerate}
	\item Draw the first center $c_1$ uniformly at random from $\pointset{Y}$
  \item Draw a new center $c_i$ by choosing $y \in \pointset{Y}$ with probability
  \begin{equation*}
    \frac{\min_{c \in \{ c^{(0)}_1, \ldots, c^{(0)}_{i-1} \}} \dist(y, c)}{
      \sum_{\tilde{y} \in \pointset{Y}} \min_{c \in \{ c^{(0)}_1, \ldots, c^{(0)}_{i-1} \}} \dist(\tilde{y}, c)}
  \end{equation*}
  \item Repeat step 2, until we have drawn $k$ centers $\pointset{A}^{(0)} = \{ c_1, \ldots, c_k \}$
  \item Assign each point $y_i \in \pointset{Y}$ to the neareast center in $\pointset{A}^{(\nu)}$
  \item Compute an estimate $c^{(\nu)}_i$ of the 1--Median of each cluster by applying (\ref{eq:Weiszfeld}) as described above and set $\pointset{A}^{(\nu+1)} := \{c^{(\nu)}_1, \ldots, c^{(\nu)}_k\}$
  \item Repeat steps 4 and 5, until no point is assigned to a new center or 10 iterations are completed
\end{enumerate}

\subsection{Construction of the final coreset}

After constructing $\pointset{A}$, we partition $\pointset{Y}$ into $\pointset{Y}_{\ell,h,a}$ as in algorithm \ref{alg:Coreset}.
From each partition $\pointset{Y}_{\ell,h,a}$, we sample $s$ points $\pointset{U}_{\ell,h,a}$ with replacement, where
\begin{equation*}
  s := \max \left( \frac{200 \cdot k}{| \{ y \in \pointset{Y}_{\ell,h,a} \, : \, |y| > 0 \} |} \, , \; 1 \right).
\end{equation*}
As mentioned before, we drop Step 5 of the algorithm by \cite{LSS12} and 
 do not compute coresets for the nodes in $\pointset{U}_{\ell,h,a}$.
The union of all $\pointset{U}_{\ell,h,a}$ constitutes the final coreset $\pointset{U}$ of $\pointset{V}$.
Algorithm \ref{alg:TurboCoreset} outlines the changes in the original steps of algorithm \ref{alg:Coreset}.

\begin{algorithm}
  \caption{\turboprobi}
  \label{alg:TurboCoreset}
  \begin{algorithmic}[\apln]
    \Function{computeCoreSet}{$\pointset{X},\pointset{V},\weight,k,\epsilon$}
      \ForAll{$v_i \in \pointset{V}$} \Comment Construct $\pointset{Y}$
        \State $y_i \gets \method{Weiszfeld::approximateOneMedian}(v_i)$
        \If{\text{iteration failed}}
          \State $v'_i \gets \emptyset$ \label{alg:tcs_gewichtung0}
          \ForAll{$x_j : p_{ij} > 0$}
            \State $v'_i \gets v'_i \cup \{x_j \, | \, n \in \left[ \, \lfloor w(v_i) p_{ij} / (w_{min} p_{min} \epsilon) \rfloor \, \right] \}$
          \EndFor \label{alg:tcs_gewichtung1}
          \State $y_i \gets \method{KumarMedian::approximateOneMedian}(v'_i, 0.9)$
        \EndIf
        \State $\weight(y_i) \gets \weight(v_i)$
      \EndFor
      \State $\pointset{Y} \gets \bigcup y_i$
      
      \State $\pointset{A} \gets \method{LloydMedian::computeCenterSet}(\pointset{Y},k)$ \Comment Construct $\pointset{A}$
        
      \State See algorithm \ref{alg:Coreset} for partitioning of \pointset{Y}
      
      \State $\pointset{U}_{\ell,h,a} \gets \method{weightedSamplingWithRep}(\pointset{V}_{\ell,h,a}, s)$
      \State $\pointset{U} \gets \bigcup_{\ell,h,a}{\pointset{U}_{\ell,h,a}}$      
      \State \Return $\pointset{U}$
    \EndFunction
  \end{algorithmic}
\end{algorithm}

\subsection{A clustering algorithm for the probabilistic k--Median problem}\label{plloyd}

In principle, every algorithm for the Euclidean probabilistic $k$-median problem can be used to compute a solution on the coreset computed by PROBI.
However, there is no standard algorithm for solving the probabilistic Euclidean $k$-median problem so far.
We use similar modifications of Lloyd's algorithm and $k$-means++ as described in Section~\ref{sec:LloydMedian} for the case of deterministic $k$-median clustering. This yields an algorithm we name \plloyd\ which we use for computing the actual solution.

Since centers are points, not nodes, we choose the first center randomly from $\pointset{X}$ (to be exact, from $\{ x_j \in \pointset{X} \, | \, \exists \, w_{ij} > 0\}$).
All remaining points $x_j$ are chosen as centers according to their \emph{total realization score} $\sum_{v_i \in \pointset{V}} w_{ij}$ and their distance to the nearest of the previously chosen centers.

In the probabilistic k--Median problem we consider in this paper, nodes are assigned to a center as a whole, i.e. independent from their actual location.
When partitioning the nodes, we assign each node $v_i$ to its expected nearest center.
The new center of a partition is constructed by computing the 1--Median of the realizations of all nodes in this partition.
We obtain the following algorithm which we will call \plloyd\ in the remainder of the report.

\begin{enumerate}
	\item Draw the first center $c_1$ uniformly at random from $\pointset{X}$
  \item Draw a new center $c_i$ by choosing $x_j \in \pointset{X}$ with probability
  \begin{equation*}
    \frac{\sum_{v_i \in \pointset{V}} \min_{c \in \{ c^{(0)}_1, \ldots, c^{(0)}_{i-1} \}} w_{ij} \; \dist(x_j, c)}{
      \sum_{\tilde{x_\ell} \in \pointset{X}} \sum_{v_i \in \pointset{V}} \min_{c \in \{ c^{(0)}_1, \ldots, c^{(0)}_{i-1} \}} w_{i\ell} \; \dist(\tilde{x_\ell}, c)}
  \end{equation*}
  \item Repeat step 2, until we have drawn $k$ centers $\pointset{C}^{(0)} = \{ c_1, \ldots, c_k \}$
  \item Assign node $v_i \in \pointset{V}$ to the expected neareast center in $\pointset{C}^{(\nu)}$
  \item Compute an estimate $c^{(\nu)}_i$ of the 1--Median of each cluster by applying (\ref{eq:Weiszfeld}) to all realizations in $c^{\nu}_i$ and set $\pointset{C}^{(\nu+1)} := \{c^{(\nu)}_1, \ldots, c^{(\nu)}_k\}$
  \item Repeat steps 4 and 5, until no node is assigned to a new center or 10 iterations are completed
\end{enumerate}

\section{Experiments}\label{sec:Experimente}

\subsection{Settings and Data} 

\paragraph{Setting.}
All computations were performed on seven machines with the same hardware configuration (2.8~Ghz Intel E7400 with 3~MB L2 Cache and 8~GB main memory).
\turboprobi{} was implemented in C++ and compiled with gcc 4.7.3.

\paragraph{Datasets.}
For the experiments, we need data sets that are large enough to test the ability of \turboprobi{} to process large amounts of data.
Furthermore, as there are no data stream implementations of algorithms for the Euclidean $k$-median that we can compare \turboprobi{} with, we wanted to at least be able to compare the speed with implementations for similar deterministic algorithms. So we decided to create probabilistic data from the data sets used in the evaluation of BICO, which is an algorithm for the deterministic $k$-means problem. Of course the problems are different (which is particularly striking when comparing the situation for $1$-median with the existence of an explicit formula for $1$-means), but in this way we at least have an indicator how long or short the time is that \turboprobi{} needs to process the data.

All points in a data set form the set $\pointset{X} = \{x_1,\ldots,x_m\}$. 
We scan the data set once to form the set of nodes $\pointset{V} = \{v_1,\ldots,v_n\}$.
Every time a chunk of 10 points $\{x_j,\ldots,x_{j+9}\}$ was read, node $v_i$ is created with a uniform distribution over its 10 realization posititions.
Overall, $\lfloor m/10 \rfloor$ nodes are constructed.
Up to 9 points at the end of the data file are omitted.

\begin{table}%
\begin{centering}
  \begin{tabular}{l||c|c|c|c|c}
                         & 	BigCross	& CalTech128 & Census & CoverType & Tower\\\hline\hline
  Number of Points       & 11620300   & 3168383    & 2458285   & 581012   & 4915200  \\ \hline
  Dimension              & 57 				& 128				 & 68 	     & 55       & 3        \\ \hline \hline
  Number of Nodes        & 1162030    & 316838     & 245828    & 58101    & 491520 \\ 
  \end{tabular}
  \caption{Sizes and dimenions of all data sets used for the experiments.\label{tab:datasets}}  
\end{centering}
\end{table}

We use five data sets. \emph{Tower}, \emph{Covertype} and \emph{Census} are data sets provided by the UCI Machine Learning Repository~\cite{BacUCI2013}.
\emph{BigCross} is a subset of the Cartesian product of \emph{Tower} and \emph{Covertype}, created by the authors of \cite{AMRSLS12} to have a very large data set.
Additionally, we used a data set we call CalTech128 which is also large and has higher dimension. 
It consists of 128 SIFT descriptors~\cite{LowDis2004} computed on the Caltech101 object database.
Table \vref{tab:datasets} gives the data set sizes and dimensions as well as the resulting number of nodes $|\pointset{V}|$.

\paragraph{Experiments.}
We ran PROBI on all five data sets with different values for $k$ which we chose similar to the values studied for BICO. However, as our point set is smaller (because every point consists of ten possible realization positions), we did not test the values of $k$ where the size of the summary would have been too large compared to the set of the original point set.

PROBI computes a solution for the Euclidean probabilistic $k$-median problem by first computing a summary and then using the adapted Lloyd algorithm. In principle, we can compare the quality of this solution with the quality of an optimal solution. 
However, due to the lack of an approximation algorithm with guaranteed quality, we have no optimal or near optimal solution to compare with.
Instead, we used \plloyd\ as described in Section~\ref{plloyd}, ran it on the whole input set to compute a solution and compared the solution quality of PROBI with the quality of this solution on Census and CoverType. 

\plloyd\ (at least in its intuitive implementation) is too memory consuming to compute solutions for the three larger data sets, so the tests on Census and CoverType are the main source of information when evaluating the quality of the solutions computed of PROBI. 
The larger data sets are still valuable to judge the speed of the algorithm. As an indication on the quality of the solutions computed for these data sets, we compared the cost of the solution on the coreset with the cost of the solution on the whole data set. This does not necessarily tell whether the solution is good (because a better solution could have a higher cost on the coreset and is thus ignored) but at least it states how accurate the summary predicts the cost for this solution.

The experiments were repeated 100 times. Tables~\ref{table:all:mean} contains the mean values, Table~\ref{table:all:median} contains the median values and Table~\ref{table:all:var} contains all variance coefficients (all three in the appendix).

\paragraph*{Quality.} 

Due to the heuristic changes, we do not expect that PROBI computes an approximation in the sense that an optimal solution on our weighted summary is always within a $(1+\varepsilon)$-fraction of the optimal solution of the point set. However, we do expect that the cost is within a constant factor, and the experiments support this idea. 
\begin{figure}[bht]
  \includegraphics[width=0.48\textwidth]{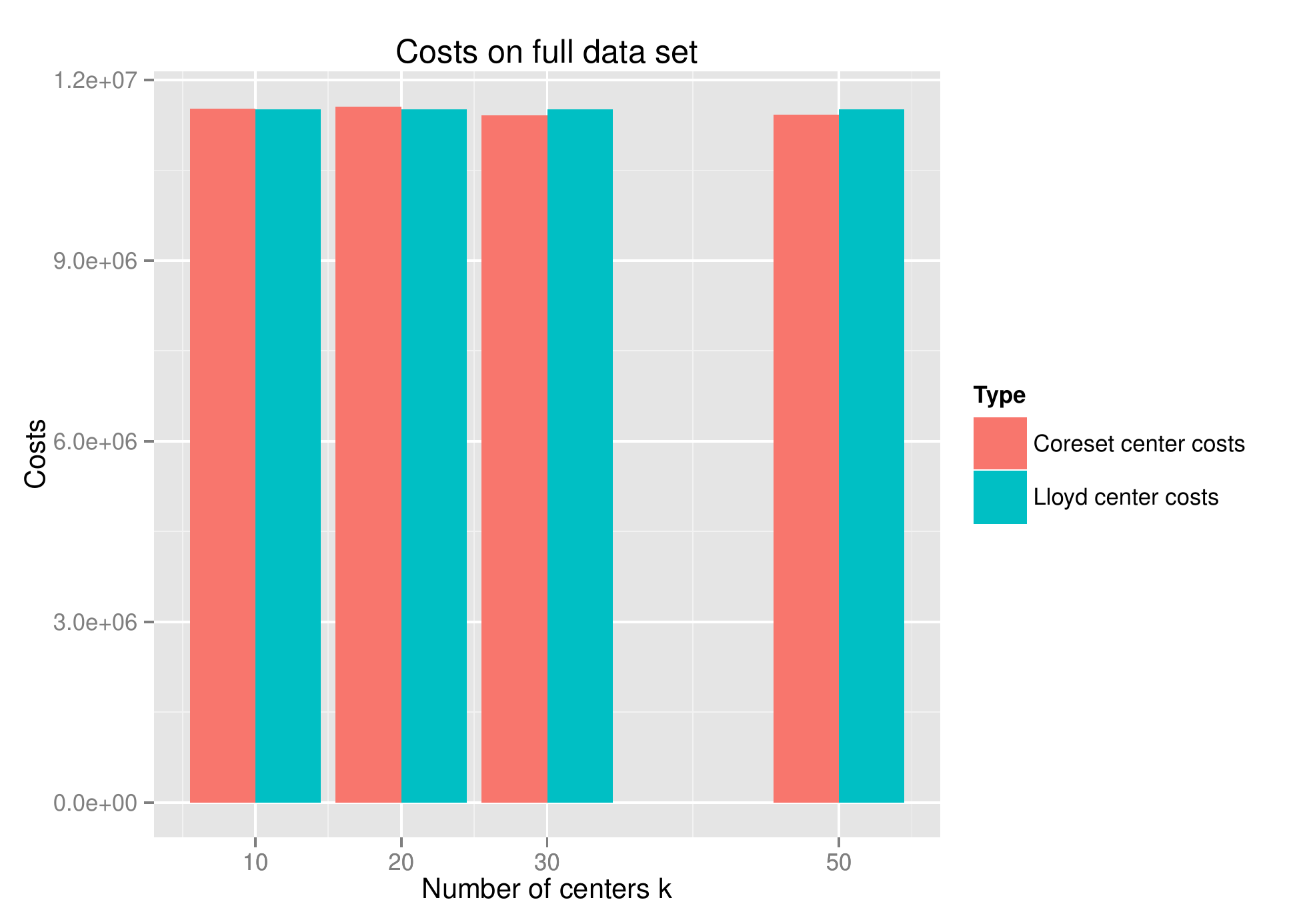}
\
  \includegraphics[width=0.48\textwidth]{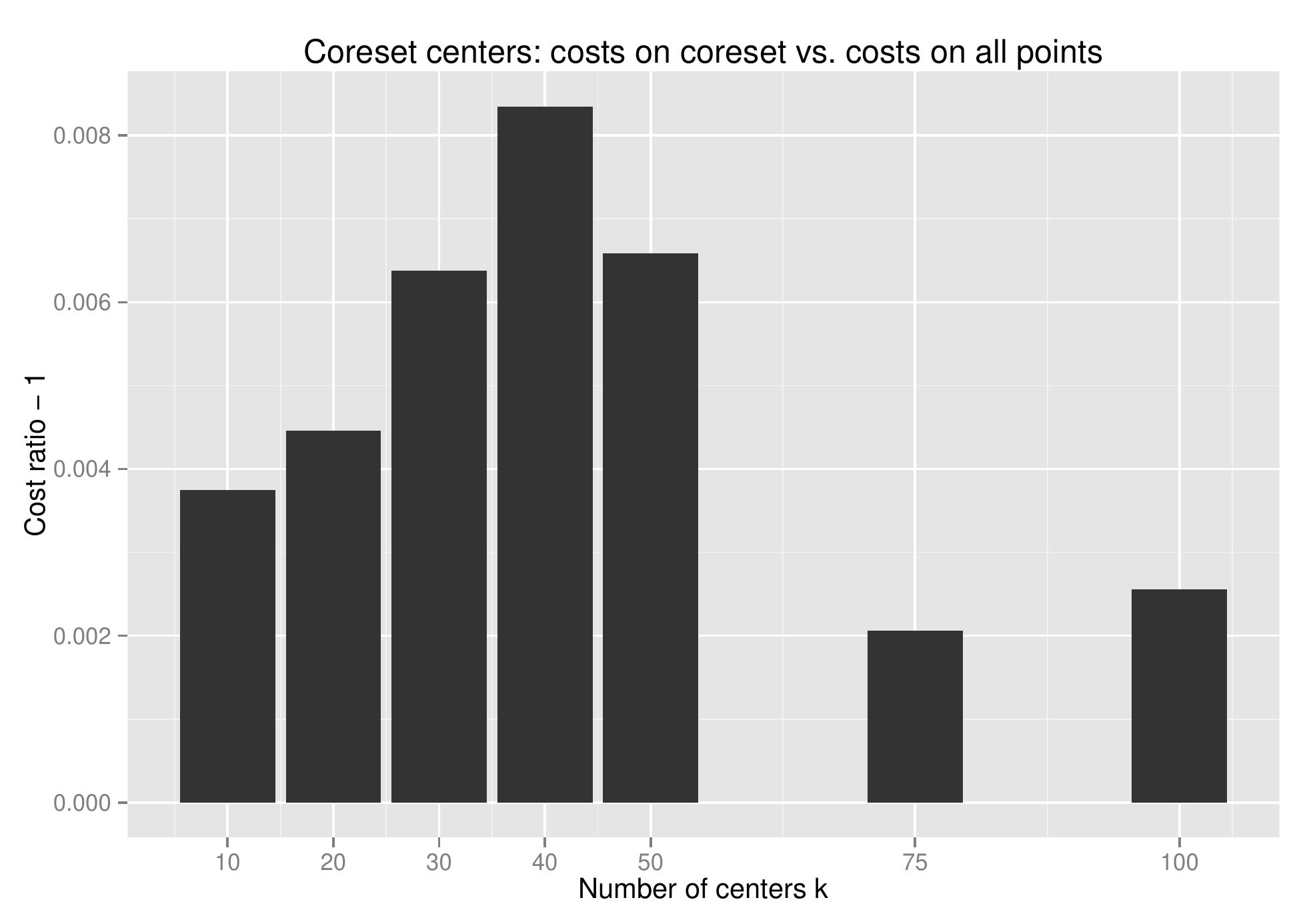}
\caption{Census Costs. Left: Cost of PROBI solution compared with \plloyd\ solution (evaluated on the full set). Right: Probi solution evaluated on summary and full data set, depicted is the difference divided by the cost on the full set.\label{diagramme:census}}
\end{figure}

\begin{figure}[thb]
  \includegraphics[width=0.48\textwidth]{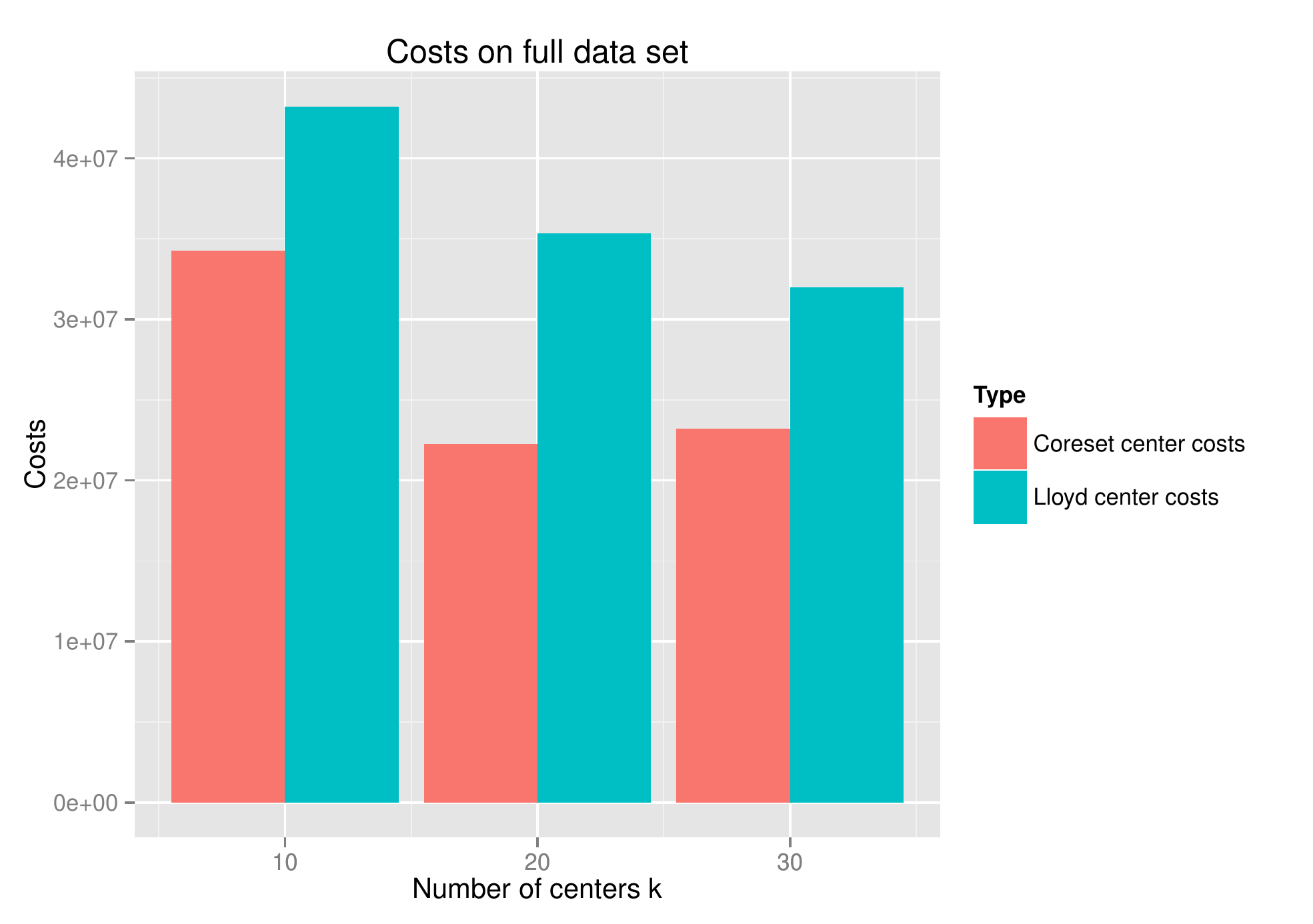}
\
  \includegraphics[width=0.48\textwidth]{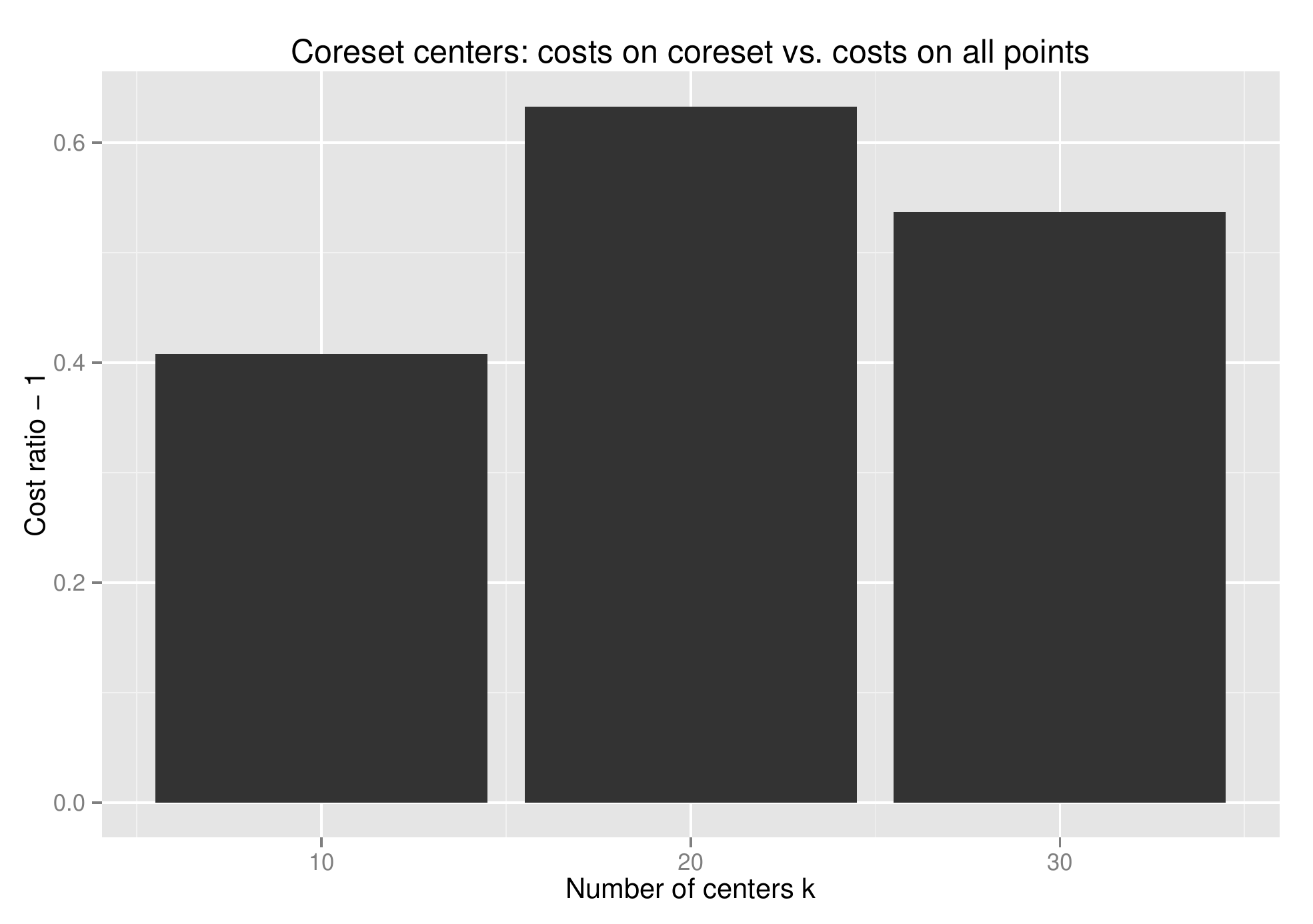}
\caption{CoverType Costs. Diagramm explanation see Figure~\ref{diagramme:census}.\label{diagramme:covtype}}
\end{figure}

For Census, the results are excellent. For all tested $k$, the cost of the solution computed by PROBI is within one percent of the cost computed by \plloyd\ on the full data set (see the left diagram in Figure~\ref{diagramme:census} and Tables~\ref{table:all:mean} and \ref{table:all:median}). For CoverType, the costs of the PROBI solutions are within two times of the cost, a little better for some $k$ (see the left diagram in Figure~\ref{diagramme:covtype}). 

It is interesting to notice that the comparison of the cost of the solution on the summary and the whole point set shows a similar behaviour (see the right diagrams in Figure~\ref{diagramme:census} and Figure~\ref{diagramme:covtype}). On Census, the difference is below one percent of the cost, on CoverType it is less than the cost (so the factor between the two is at most two). In particular, the difference is largest for $k=20$, which is indeed the case where the solution quality is worst, and the quantities also look connected. This indicates that looking at how much the cost of the PROBI solution changes when evaluating it on the whole point set instead of the summary gives a hint on the solution quality.

\begin{figure}[b]
  \includegraphics[width=0.48\textwidth]{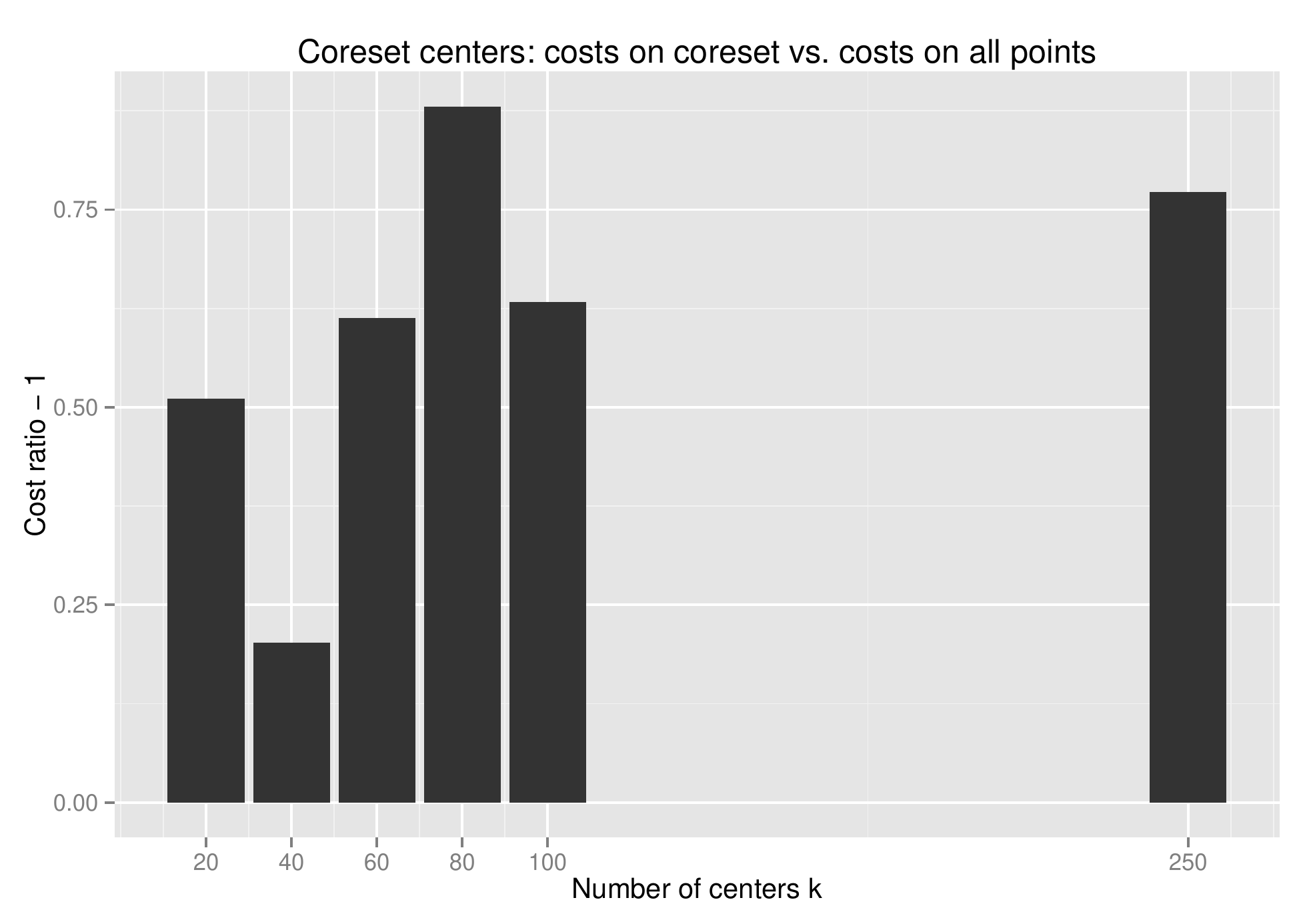}
\
  \includegraphics[width=0.48\textwidth]{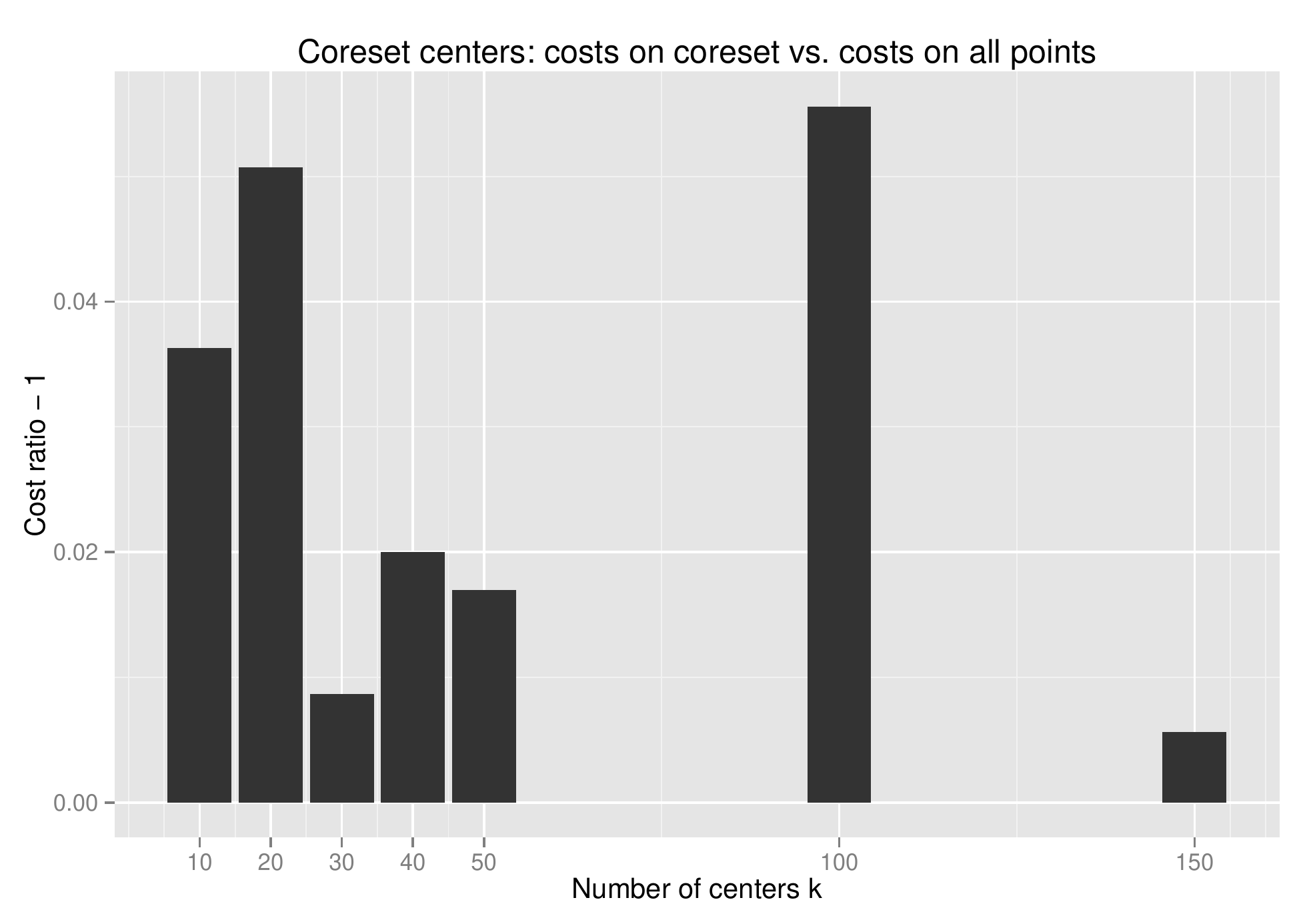}
\caption{Tower and CalTech Costs. Diagramm explanations see Figure~\ref{diagramme:bigcrosscosts}.\label{diagramme:towercaltechcosts}}
\end{figure}

\begin{figure}[t]
  \includegraphics[width=\textwidth, height=8cm]{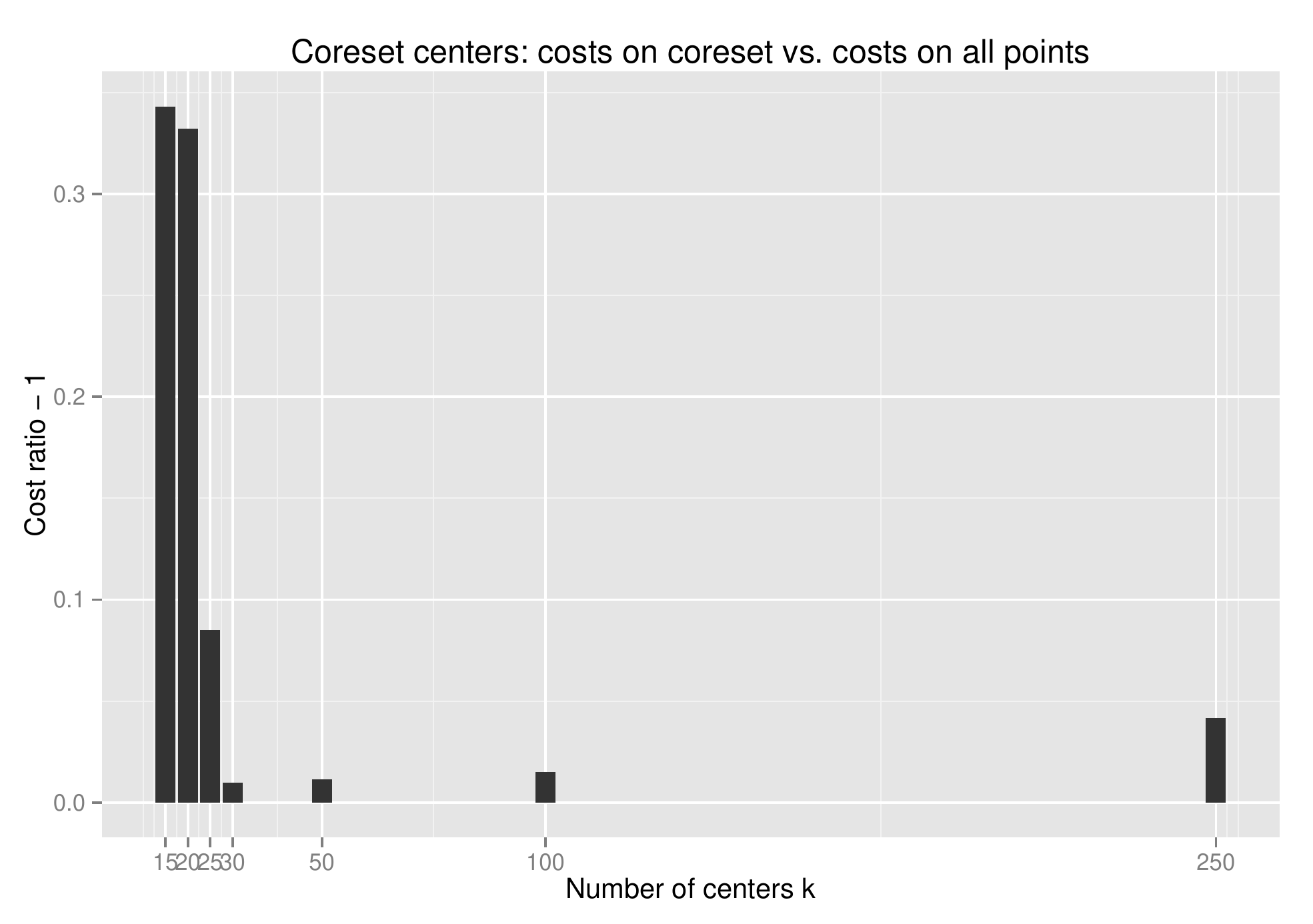}
\caption{BigCross Costs. Diagramm depicts difference between evaluation on summary and full set, normalized by dividing by the cost on the full set.\label{diagramme:bigcrosscosts}}
\end{figure}

Based on this, the results for BigCross (Figure~\ref{diagramme:bigcrosscosts}), Tower (left diagramm in Figure~\ref{diagramme:towercaltechcosts}) and Caltech (right diagramm in Figure~\ref{diagramme:towercaltechcosts}) indicate that the solution quality might be within a factor of two as well (because the difference in the cost divided by the cost is always less than 1).

\paragraph*{Running time.} 
Figures~\ref{diagramme:bigcrosstime}, \ref{diagramme:cctime} and~\ref{diagramme:towercaltechtime} show the running times of PROBI for all tested data sets. The time increases linearly with the number of centers and we suspect that this is due to the computation of a bicriteria approximation using \plloyd, which has a relatively high running time. 

Again, there are no obvious algorithms to compare the running time of PROBI with. PROBI is much faster than the adapted $k$-means++ algorithm \plloyd, which is not surprising in light of similar results for StreamKM++ compared with $k$-means++ in \cite{AMRSLS12}. The experiments for PROBI were performed on the same machines as those for BICO in \cite{FGSSS13}. This allows us to compare the running times. Comparing with Figures 2, 3 and 4 in \cite{FGSSS13}, we see that the worst case of the ratio between PROBIs and BICOs 
running time is around 2 considering all cases where both algorithms were tested. This 
makes PROBIs running time comparable to BICO and much faster than for example StreamKM++. Notice however that PROBIs running time increases faster than BICOs with increasing $k$ (but linearly).

\begin{Bild}
  \includegraphics[width=\textwidth, height=8cm]{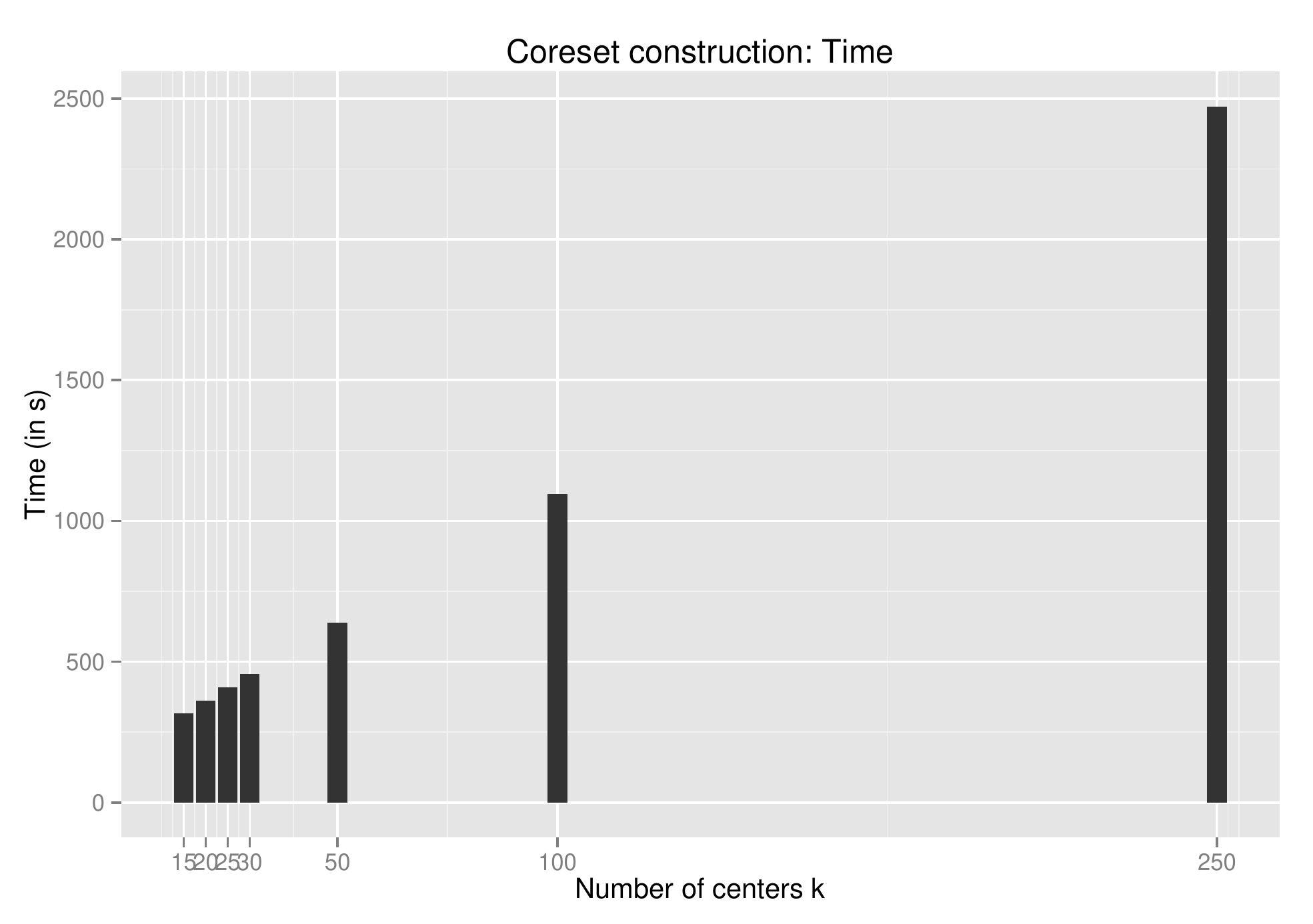}
\captionof{figure}{BigCross Running Times.\label{diagramme:bigcrosstime}}
\end{Bild}

\begin{Bild}
  \includegraphics[width=0.43\textwidth]{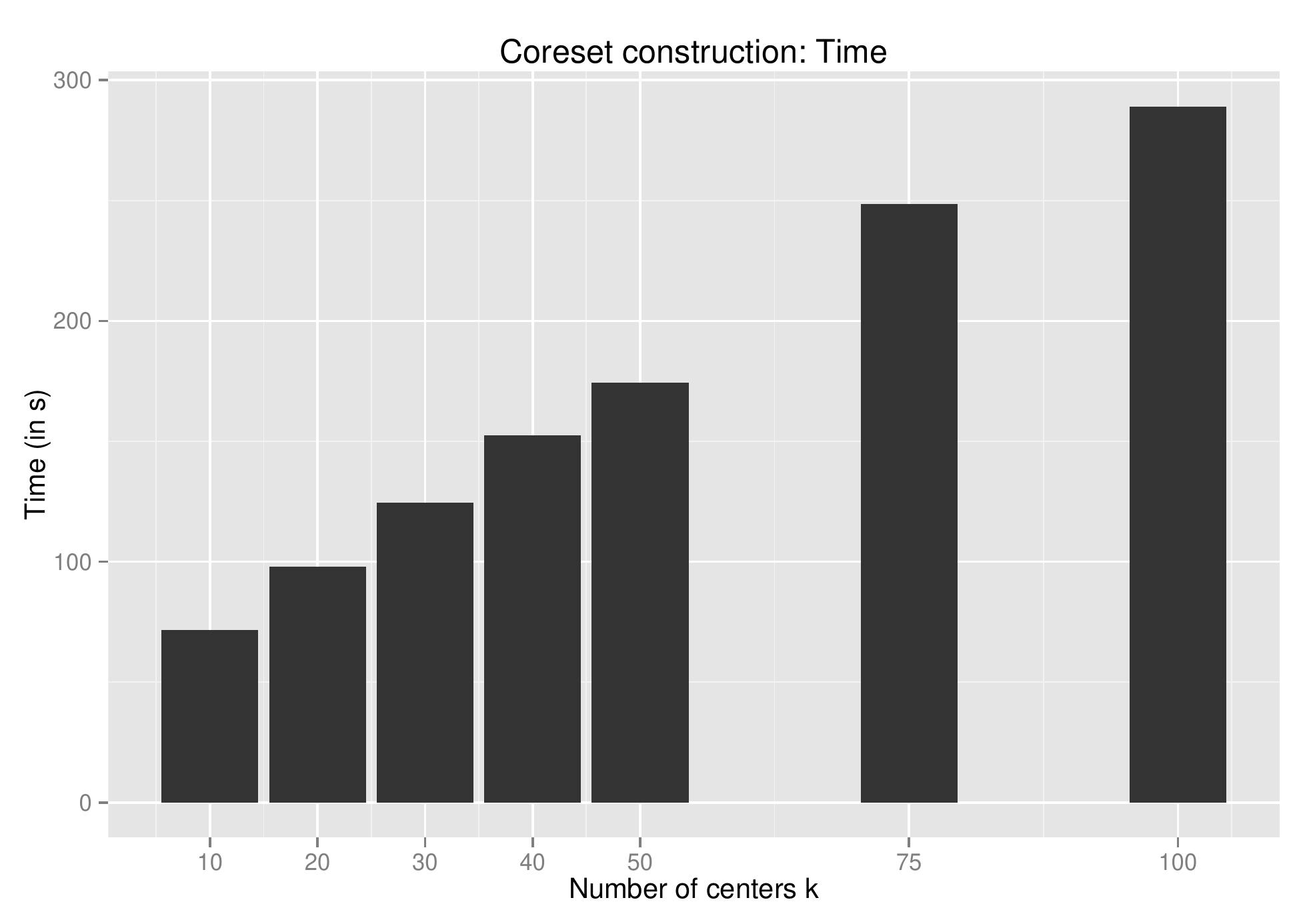}
\
  \includegraphics[width=0.43\textwidth]{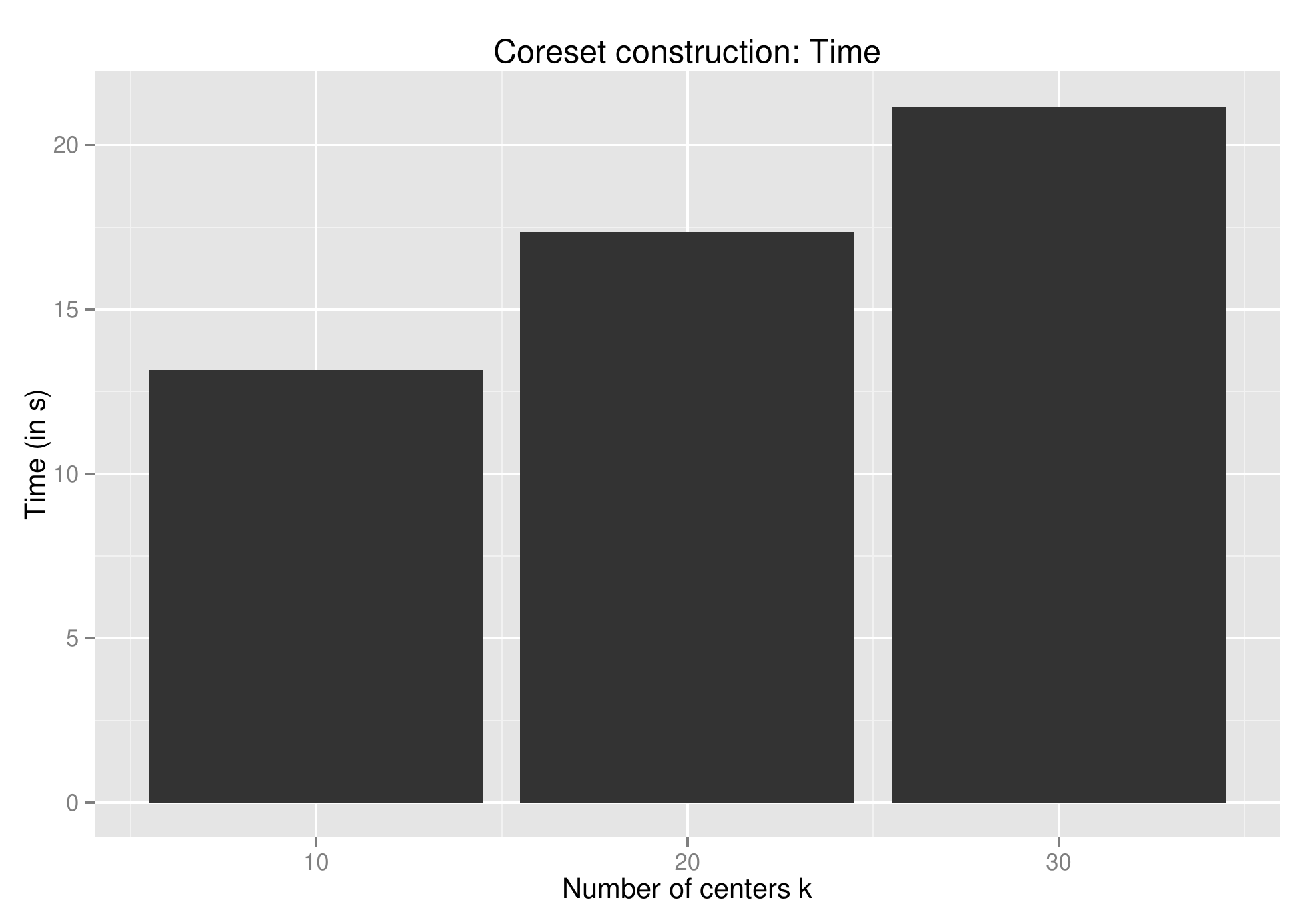}
\captionof{figure}{Census and CoverType Running Times.\label{diagramme:cctime}}
\end{Bild}
\vspace*{-0.5cm}
\begin{Bild}
   \includegraphics[width=0.43\textwidth]{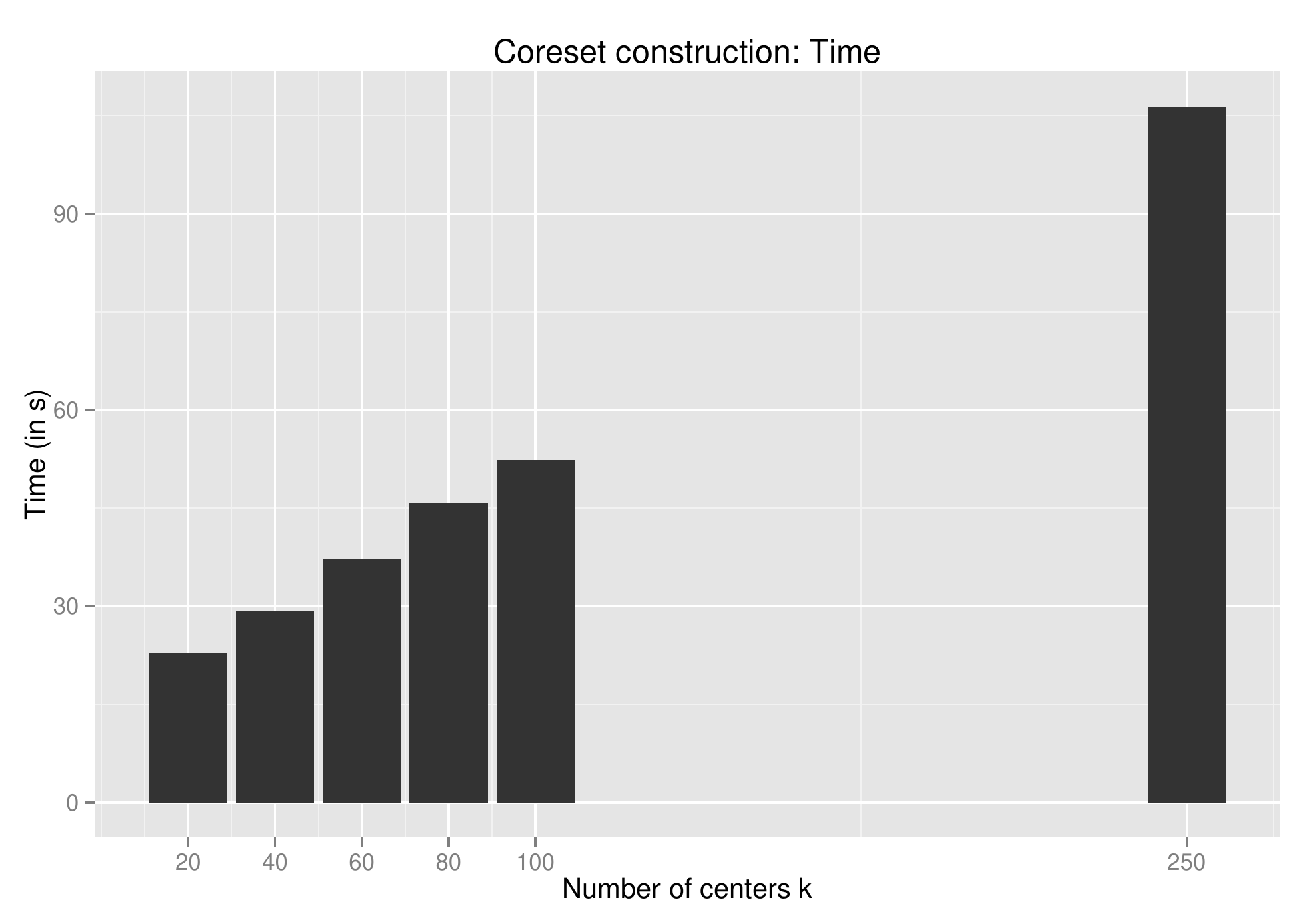}
\
   \includegraphics[width=0.43\textwidth]{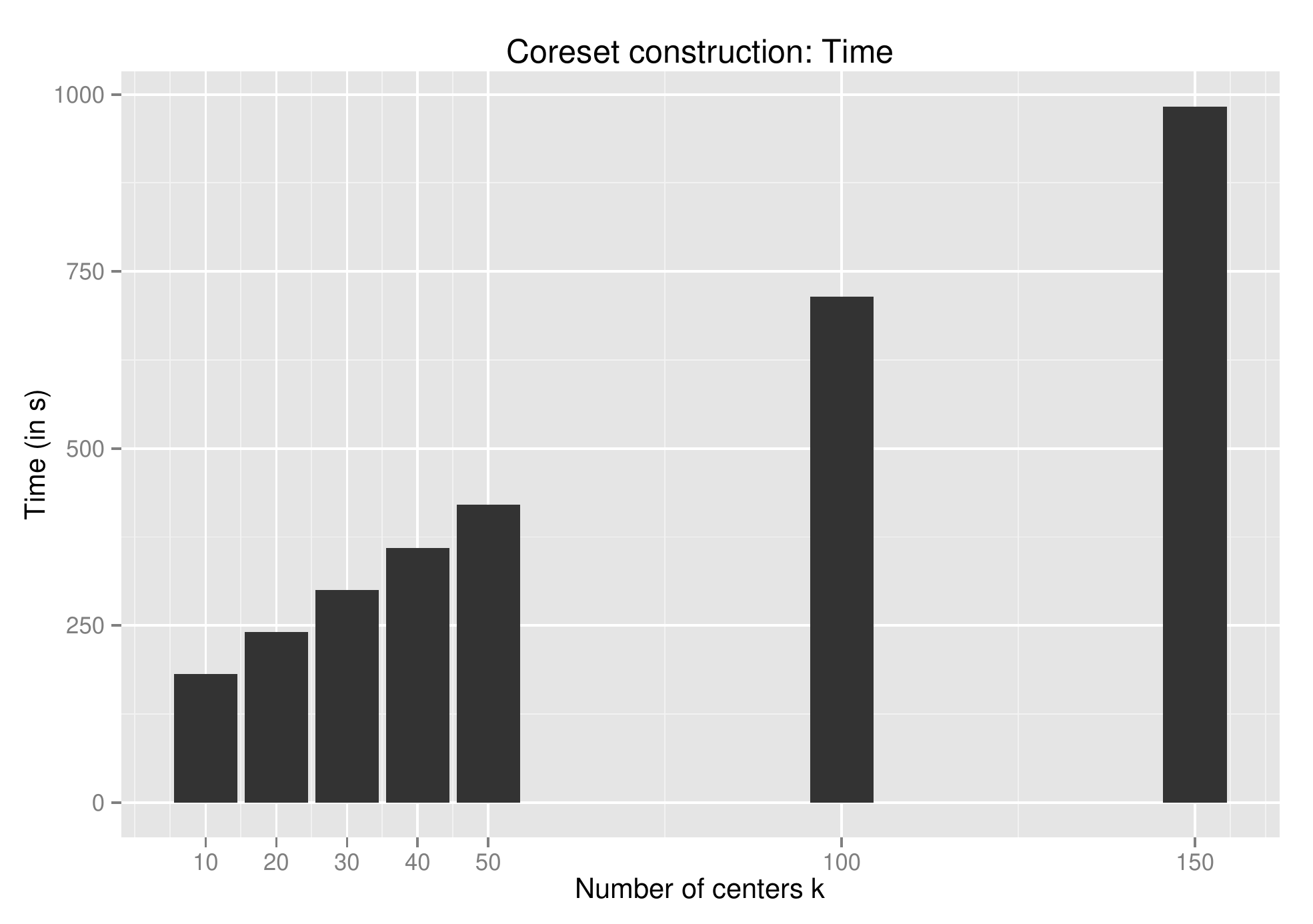}
\captionof{figure}{Tower and CalTech Running Times.\label{diagramme:towercaltechtime}}
\end{Bild}

\paragraph*{Acknowledgements}
This work has been supported by Deutsche Forschungsgemeinschaft (DFG) within grant SO 514/4-3. 
The project CG Learning acknowledges the financial support of the Future and Emerging Technologies (FET) program within the Seventh Framework Programme for Research of the European Commission, under FET-Open grant number: 255827

\newpage

\bibliographystyle{abbrv}
\bibliography{Literatur}

\appendix

\begin{table}%
\begin{centering}
  \begin{tabular}{|l||c|c|c|c|c|c|c|c|}
\hline
           &     & \multicolumn{3}{|c|}{Costs}                                                                                                       & \multicolumn{3}{|c|}{Running Time}                                                                                               \\ \hline
   Dataset &   k & \multicolumn{2}{|c|}{PROBI}                                                           &                                 P-LLOYD++ & \multicolumn{2}{|c|}{PROBI}                                                           &                                 P-LLOYD++\\ \hline \hline
  BigCross &  15 & $7.55\text{\textbf e} \boldsymbol{ +08 }$ & $1.15\text{\textbf e} \boldsymbol{ +09 }$ &                                       --- & $3.16\text{\textbf e} \boldsymbol{ +02 }$ & $3.82\text{\textbf e} \boldsymbol{ +01 }$ &                                       ---\\ \hline
           &  20 & $7.92\text{\textbf e} \boldsymbol{ +08 }$ & $1.19\text{\textbf e} \boldsymbol{ +09 }$ &                                       --- & $3.62\text{\textbf e} \boldsymbol{ +02 }$ & $8.45\text{\textbf e} \boldsymbol{ +01 }$ &                                       ---\\ \hline
           &  25 & $8.28\text{\textbf e} \boldsymbol{ +08 }$ & $9.05\text{\textbf e} \boldsymbol{ +08 }$ &                                       --- & $4.09\text{\textbf e} \boldsymbol{ +02 }$ & $1.73\text{\textbf e} \boldsymbol{ +02 }$ &                                       ---\\ \hline
           &  30 & $9.40\text{\textbf e} \boldsymbol{ +08 }$ & $9.44\text{\textbf e} \boldsymbol{ +08 }$ &                                       --- & $4.57\text{\textbf e} \boldsymbol{ +02 }$ & $2.42\text{\textbf e} \boldsymbol{ +02 }$ &                                       ---\\ \hline
           &  50 & $8.38\text{\textbf e} \boldsymbol{ +08 }$ & $8.30\text{\textbf e} \boldsymbol{ +08 }$ &                                       --- & $6.39\text{\textbf e} \boldsymbol{ +02 }$ & $8.54\text{\textbf e} \boldsymbol{ +02 }$ &                                       ---\\ \hline
           & 100 & $7.98\text{\textbf e} \boldsymbol{ +08 }$ & $7.87\text{\textbf e} \boldsymbol{ +08 }$ &                                       --- & $1.10\text{\textbf e} \boldsymbol{ +03 }$ & $2.82\text{\textbf e} \boldsymbol{ +03 }$ &                                       ---\\ \hline
           & 250 & $7.21\text{\textbf e} \boldsymbol{ +08 }$ & $7.52\text{\textbf e} \boldsymbol{ +08 }$ &                                       --- & $2.47\text{\textbf e} \boldsymbol{ +03 }$ & $8.49\text{\textbf e} \boldsymbol{ +03 }$ &                                       ---\\ \hline \hline
Caltech128 &  10 & $1.15\text{\textbf e} \boldsymbol{ +08 }$ & $1.19\text{\textbf e} \boldsymbol{ +08 }$ &                                       --- & $1.81\text{\textbf e} \boldsymbol{ +02 }$ & $2.22\text{\textbf e} \boldsymbol{ +01 }$ &                                       ---\\ \hline
           &  20 & $1.11\text{\textbf e} \boldsymbol{ +08 }$ & $1.16\text{\textbf e} \boldsymbol{ +08 }$ &                                       --- & $2.41\text{\textbf e} \boldsymbol{ +02 }$ & $1.28\text{\textbf e} \boldsymbol{ +02 }$ &                                       ---\\ \hline
           &  30 & $1.14\text{\textbf e} \boldsymbol{ +08 }$ & $1.15\text{\textbf e} \boldsymbol{ +08 }$ &                                       --- & $3.00\text{\textbf e} \boldsymbol{ +02 }$ & $8.51\text{\textbf e} \boldsymbol{ +02 }$ &                                       ---\\ \hline
           &  40 & $1.12\text{\textbf e} \boldsymbol{ +08 }$ & $1.14\text{\textbf e} \boldsymbol{ +08 }$ &                                       --- & $3.59\text{\textbf e} \boldsymbol{ +02 }$ & $4.41\text{\textbf e} \boldsymbol{ +02 }$ &                                       ---\\ \hline
           &  50 & $1.13\text{\textbf e} \boldsymbol{ +08 }$ & $1.15\text{\textbf e} \boldsymbol{ +08 }$ &                                       --- & $4.20\text{\textbf e} \boldsymbol{ +02 }$ & $2.10\text{\textbf e} \boldsymbol{ +02 }$ &                                       ---\\ \hline
           & 100 & $1.07\text{\textbf e} \boldsymbol{ +08 }$ & $1.13\text{\textbf e} \boldsymbol{ +08 }$ &                                       --- & $7.14\text{\textbf e} \boldsymbol{ +02 }$ & $1.37\text{\textbf e} \boldsymbol{ +03 }$ &                                       ---\\ \hline
           & 150 & $1.12\text{\textbf e} \boldsymbol{ +08 }$ & $1.11\text{\textbf e} \boldsymbol{ +08 }$ &                                       --- & $9.83\text{\textbf e} \boldsymbol{ +02 }$ & $9.76\text{\textbf e} \boldsymbol{ +03 }$ &                                       ---\\ \hline \hline
    Census &  10 & $1.15\text{\textbf e} \boldsymbol{ +07 }$ & $1.15\text{\textbf e} \boldsymbol{ +07 }$ & $1.15\text{\textbf e} \boldsymbol{ +07 }$ & $7.16\text{\textbf e} \boldsymbol{ +01 }$ & $5.94\text{\textbf e} \boldsymbol{ +01 }$ & $6.97\text{\textbf e} \boldsymbol{ +03 }$\\ \hline
           &  20 & $1.16\text{\textbf e} \boldsymbol{ +07 }$ & $1.15\text{\textbf e} \boldsymbol{ +07 }$ & $1.15\text{\textbf e} \boldsymbol{ +07 }$ & $9.79\text{\textbf e} \boldsymbol{ +01 }$ & $2.83\text{\textbf e} \boldsymbol{ +02 }$ & $1.09\text{\textbf e} \boldsymbol{ +04 }$\\ \hline
           &  30 & $1.14\text{\textbf e} \boldsymbol{ +07 }$ & $1.15\text{\textbf e} \boldsymbol{ +07 }$ & $1.15\text{\textbf e} \boldsymbol{ +07 }$ & $1.25\text{\textbf e} \boldsymbol{ +02 }$ & $4.34\text{\textbf e} \boldsymbol{ +02 }$ & $1.49\text{\textbf e} \boldsymbol{ +04 }$\\ \hline
           &  40 & $1.14\text{\textbf e} \boldsymbol{ +07 }$ & $1.15\text{\textbf e} \boldsymbol{ +07 }$ &                                       --- & $1.52\text{\textbf e} \boldsymbol{ +02 }$ & $2.67\text{\textbf e} \boldsymbol{ +02 }$ &                                       ---\\ \hline
           &  50 & $1.14\text{\textbf e} \boldsymbol{ +07 }$ & $1.15\text{\textbf e} \boldsymbol{ +07 }$ & $1.15\text{\textbf e} \boldsymbol{ +07 }$ & $1.74\text{\textbf e} \boldsymbol{ +02 }$ & $1.23\text{\textbf e} \boldsymbol{ +03 }$ & $2.29\text{\textbf e} \boldsymbol{ +04 }$\\ \hline
           &  75 & $1.15\text{\textbf e} \boldsymbol{ +07 }$ & $1.15\text{\textbf e} \boldsymbol{ +07 }$ &                                       --- & $2.49\text{\textbf e} \boldsymbol{ +02 }$ & $1.64\text{\textbf e} \boldsymbol{ +03 }$ &                                       ---\\ \hline
           & 100 & $1.15\text{\textbf e} \boldsymbol{ +07 }$ & $1.15\text{\textbf e} \boldsymbol{ +07 }$ &                                       --- & $2.89\text{\textbf e} \boldsymbol{ +02 }$ & $4.28\text{\textbf e} \boldsymbol{ +03 }$ &                                       ---\\ \hline \hline
 Covertype &  10 & $3.42\text{\textbf e} \boldsymbol{ +07 }$ & $5.79\text{\textbf e} \boldsymbol{ +07 }$ & $4.32\text{\textbf e} \boldsymbol{ +07 }$ & $1.31\text{\textbf e} \boldsymbol{ +01 }$ & $1.81\text{\textbf e} \boldsymbol{ +01 }$ & $1.01\text{\textbf e} \boldsymbol{ +03 }$\\ \hline
           &  20 & $2.23\text{\textbf e} \boldsymbol{ +07 }$ & $6.06\text{\textbf e} \boldsymbol{ +07 }$ & $3.53\text{\textbf e} \boldsymbol{ +07 }$ & $1.74\text{\textbf e} \boldsymbol{ +01 }$ & $1.42\text{\textbf e} \boldsymbol{ +01 }$ & $2.19\text{\textbf e} \boldsymbol{ +03 }$\\ \hline
           &  30 & $2.32\text{\textbf e} \boldsymbol{ +07 }$ & $5.02\text{\textbf e} \boldsymbol{ +07 }$ & $3.20\text{\textbf e} \boldsymbol{ +07 }$ & $2.12\text{\textbf e} \boldsymbol{ +01 }$ & $1.47\text{\textbf e} \boldsymbol{ +02 }$ & $2.97\text{\textbf e} \boldsymbol{ +03 }$\\ \hline \hline
     Tower &  20 & $5.08\text{\textbf e} \boldsymbol{ +06 }$ & $1.05\text{\textbf e} \boldsymbol{ +07 }$ &                                       --- & $2.28\text{\textbf e} \boldsymbol{ +01 }$ & $1.55\text{\textbf e} \boldsymbol{ +01 }$ &                                       ---\\ \hline
           &  40 & $6.37\text{\textbf e} \boldsymbol{ +06 }$ & $7.99\text{\textbf e} \boldsymbol{ +06 }$ &                                       --- & $2.92\text{\textbf e} \boldsymbol{ +01 }$ & $5.01\text{\textbf e} \boldsymbol{ +01 }$ &                                       ---\\ \hline
           &  60 & $3.41\text{\textbf e} \boldsymbol{ +06 }$ & $8.81\text{\textbf e} \boldsymbol{ +06 }$ &                                       --- & $3.72\text{\textbf e} \boldsymbol{ +01 }$ & $6.77\text{\textbf e} \boldsymbol{ +01 }$ &                                       ---\\ \hline
           &  80 & $1.96\text{\textbf e} \boldsymbol{ +06 }$ & $1.67\text{\textbf e} \boldsymbol{ +07 }$ &                                       --- & $4.59\text{\textbf e} \boldsymbol{ +01 }$ & $2.73\text{\textbf e} \boldsymbol{ +01 }$ &                                       ---\\ \hline
           & 100 & $2.89\text{\textbf e} \boldsymbol{ +06 }$ & $7.96\text{\textbf e} \boldsymbol{ +06 }$ &                                       --- & $5.23\text{\textbf e} \boldsymbol{ +01 }$ & $1.84\text{\textbf e} \boldsymbol{ +02 }$ &                                       ---\\ \hline
           & 250 & $2.08\text{\textbf e} \boldsymbol{ +06 }$ & $9.16\text{\textbf e} \boldsymbol{ +06 }$ &                                       --- & $1.06\text{\textbf e} \boldsymbol{ +02 }$ & $1.05\text{\textbf e} \boldsymbol{ +03 }$ &                                       ---\\ \hline
\end{tabular}
 
  \caption{Mean values of 100 runs on all tested data sets and $k$. The first cost column for PROBI gives the cost on the summary, the second gives the cost on the full data set. The first PROBI column in the running time part gives the running time of PROBI, the second  indicates the additional time needed by \plloyd\ to compute the solution on the summary. The previous diagrams only show the PROBI time\label{table:all:mean}.}
\end{centering}
\end{table}
\begin{table}%
\begin{centering}
  \begin{tabular}{|l||c|c|c|c|c|c|c|}
\hline
           &     & \multicolumn{3}{|c|}{Costs}                                                                                                       & \multicolumn{3}{|c|}{Running Time}                                                                                               \\ \hline
   Dataset &   k & \multicolumn{2}{|c|}{PROBI}                                                           &                                 P-LLOYD++ & \multicolumn{2}{|c|}{PROBI}                                                           &                                 P-LLOYD++\\ \hline \hline
  BigCross &  15 & $7.55\text{\textbf e} \boldsymbol{ +08 }$ & $1.16\text{\textbf e} \boldsymbol{ +09 }$ &                                       --- & $3.15\text{\textbf e} \boldsymbol{ +02 }$ & $3.79\text{\textbf e} \boldsymbol{ +01 }$ &                                       ---\\ \hline
           &  20 & $7.92\text{\textbf e} \boldsymbol{ +08 }$ & $1.19\text{\textbf e} \boldsymbol{ +09 }$ &                                       --- & $3.62\text{\textbf e} \boldsymbol{ +02 }$ & $8.47\text{\textbf e} \boldsymbol{ +01 }$ &                                       ---\\ \hline
           &  25 & $8.28\text{\textbf e} \boldsymbol{ +08 }$ & $9.05\text{\textbf e} \boldsymbol{ +08 }$ &                                       --- & $4.08\text{\textbf e} \boldsymbol{ +02 }$ & $1.71\text{\textbf e} \boldsymbol{ +02 }$ &                                       ---\\ \hline
           &  30 & $9.40\text{\textbf e} \boldsymbol{ +08 }$ & $9.43\text{\textbf e} \boldsymbol{ +08 }$ &                                       --- & $4.57\text{\textbf e} \boldsymbol{ +02 }$ & $2.46\text{\textbf e} \boldsymbol{ +02 }$ &                                       ---\\ \hline
           &  50 & $8.36\text{\textbf e} \boldsymbol{ +08 }$ & $8.30\text{\textbf e} \boldsymbol{ +08 }$ &                                       --- & $6.38\text{\textbf e} \boldsymbol{ +02 }$ & $8.57\text{\textbf e} \boldsymbol{ +02 }$ &                                       ---\\ \hline
           & 100 & $7.98\text{\textbf e} \boldsymbol{ +08 }$ & $7.87\text{\textbf e} \boldsymbol{ +08 }$ &                                       --- & $1.10\text{\textbf e} \boldsymbol{ +03 }$ & $2.82\text{\textbf e} \boldsymbol{ +03 }$ &                                       ---\\ \hline
           & 250 & $7.21\text{\textbf e} \boldsymbol{ +08 }$ & $7.52\text{\textbf e} \boldsymbol{ +08 }$ &                                       --- & $2.45\text{\textbf e} \boldsymbol{ +03 }$ & $8.47\text{\textbf e} \boldsymbol{ +03 }$ &                                       ---\\ \hline \hline
Caltech128 &  10 & $1.15\text{\textbf e} \boldsymbol{ +08 }$ & $1.19\text{\textbf e} \boldsymbol{ +08 }$ &                                       --- & $1.81\text{\textbf e} \boldsymbol{ +02 }$ & $2.17\text{\textbf e} \boldsymbol{ +01 }$ &                                       ---\\ \hline
           &  20 & $1.11\text{\textbf e} \boldsymbol{ +08 }$ & $1.16\text{\textbf e} \boldsymbol{ +08 }$ &                                       --- & $2.41\text{\textbf e} \boldsymbol{ +02 }$ & $1.28\text{\textbf e} \boldsymbol{ +02 }$ &                                       ---\\ \hline
           &  30 & $1.14\text{\textbf e} \boldsymbol{ +08 }$ & $1.15\text{\textbf e} \boldsymbol{ +08 }$ &                                       --- & $2.99\text{\textbf e} \boldsymbol{ +02 }$ & $8.60\text{\textbf e} \boldsymbol{ +02 }$ &                                       ---\\ \hline
           &  40 & $1.12\text{\textbf e} \boldsymbol{ +08 }$ & $1.15\text{\textbf e} \boldsymbol{ +08 }$ &                                       --- & $3.59\text{\textbf e} \boldsymbol{ +02 }$ & $4.40\text{\textbf e} \boldsymbol{ +02 }$ &                                       ---\\ \hline
           &  50 & $1.13\text{\textbf e} \boldsymbol{ +08 }$ & $1.15\text{\textbf e} \boldsymbol{ +08 }$ &                                       --- & $4.20\text{\textbf e} \boldsymbol{ +02 }$ & $2.05\text{\textbf e} \boldsymbol{ +02 }$ &                                       ---\\ \hline
           & 100 & $1.07\text{\textbf e} \boldsymbol{ +08 }$ & $1.13\text{\textbf e} \boldsymbol{ +08 }$ &                                       --- & $7.14\text{\textbf e} \boldsymbol{ +02 }$ & $1.37\text{\textbf e} \boldsymbol{ +03 }$ &                                       ---\\ \hline
           & 150 & $1.12\text{\textbf e} \boldsymbol{ +08 }$ & $1.11\text{\textbf e} \boldsymbol{ +08 }$ &                                       --- & $9.83\text{\textbf e} \boldsymbol{ +02 }$ & $9.78\text{\textbf e} \boldsymbol{ +03 }$ &                                       ---\\ \hline \hline
    Census &  10 & $1.15\text{\textbf e} \boldsymbol{ +07 }$ & $1.15\text{\textbf e} \boldsymbol{ +07 }$ & $1.15\text{\textbf e} \boldsymbol{ +07 }$ & $7.15\text{\textbf e} \boldsymbol{ +01 }$ & $5.74\text{\textbf e} \boldsymbol{ +01 }$ & $6.94\text{\textbf e} \boldsymbol{ +03 }$\\ \hline
           &  20 & $1.16\text{\textbf e} \boldsymbol{ +07 }$ & $1.15\text{\textbf e} \boldsymbol{ +07 }$ & $1.15\text{\textbf e} \boldsymbol{ +07 }$ & $9.78\text{\textbf e} \boldsymbol{ +01 }$ & $2.76\text{\textbf e} \boldsymbol{ +02 }$ & $1.09\text{\textbf e} \boldsymbol{ +04 }$\\ \hline
           &  30 & $1.14\text{\textbf e} \boldsymbol{ +07 }$ & $1.15\text{\textbf e} \boldsymbol{ +07 }$ & $1.15\text{\textbf e} \boldsymbol{ +07 }$ & $1.25\text{\textbf e} \boldsymbol{ +02 }$ & $4.49\text{\textbf e} \boldsymbol{ +02 }$ & $1.48\text{\textbf e} \boldsymbol{ +04 }$\\ \hline
           &  40 & $1.14\text{\textbf e} \boldsymbol{ +07 }$ & $1.15\text{\textbf e} \boldsymbol{ +07 }$ &                                           & $1.52\text{\textbf e} \boldsymbol{ +02 }$ & $2.75\text{\textbf e} \boldsymbol{ +02 }$ &                                          \\ \hline
           &  50 & $1.14\text{\textbf e} \boldsymbol{ +07 }$ & $1.15\text{\textbf e} \boldsymbol{ +07 }$ & $1.15\text{\textbf e} \boldsymbol{ +07 }$ & $1.74\text{\textbf e} \boldsymbol{ +02 }$ & $1.22\text{\textbf e} \boldsymbol{ +03 }$ & $2.27\text{\textbf e} \boldsymbol{ +04 }$\\ \hline
           &  75 & $1.15\text{\textbf e} \boldsymbol{ +07 }$ & $1.15\text{\textbf e} \boldsymbol{ +07 }$ &                                           & $2.44\text{\textbf e} \boldsymbol{ +02 }$ & $1.64\text{\textbf e} \boldsymbol{ +03 }$ &                                          \\ \hline
           & 100 & $1.15\text{\textbf e} \boldsymbol{ +07 }$ & $1.15\text{\textbf e} \boldsymbol{ +07 }$ &                                           & $2.89\text{\textbf e} \boldsymbol{ +02 }$ & $4.28\text{\textbf e} \boldsymbol{ +03 }$ &                                          \\ \hline \hline
 Covertype &  10 & $3.43\text{\textbf e} \boldsymbol{ +07 }$ & $5.79\text{\textbf e} \boldsymbol{ +07 }$ & $4.32\text{\textbf e} \boldsymbol{ +07 }$ & $1.31\text{\textbf e} \boldsymbol{ +01 }$ & $1.72\text{\textbf e} \boldsymbol{ +01 }$ & $1.01\text{\textbf e} \boldsymbol{ +03 }$\\ \hline
           &  20 & $2.23\text{\textbf e} \boldsymbol{ +07 }$ & $6.07\text{\textbf e} \boldsymbol{ +07 }$ & $3.54\text{\textbf e} \boldsymbol{ +07 }$ & $1.73\text{\textbf e} \boldsymbol{ +01 }$ & $1.43\text{\textbf e} \boldsymbol{ +01 }$ & $2.11\text{\textbf e} \boldsymbol{ +03 }$\\ \hline
           &  30 & $2.32\text{\textbf e} \boldsymbol{ +07 }$ & $5.02\text{\textbf e} \boldsymbol{ +07 }$ & $3.20\text{\textbf e} \boldsymbol{ +07 }$ & $2.12\text{\textbf e} \boldsymbol{ +01 }$ & $1.45\text{\textbf e} \boldsymbol{ +02 }$ & $2.92\text{\textbf e} \boldsymbol{ +03 }$\\ \hline \hline
     Tower &  20 & $5.04\text{\textbf e} \boldsymbol{ +06 }$ & $1.02\text{\textbf e} \boldsymbol{ +07 }$ &                                       --- & $2.22\text{\textbf e} \boldsymbol{ +01 }$ & $1.56\text{\textbf e} \boldsymbol{ +01 }$ &                                       ---\\ \hline
           &  40 & $6.30\text{\textbf e} \boldsymbol{ +06 }$ & $7.99\text{\textbf e} \boldsymbol{ +06 }$ &                                       --- & $2.92\text{\textbf e} \boldsymbol{ +01 }$ & $5.00\text{\textbf e} \boldsymbol{ +01 }$ &                                       ---\\ \hline
           &  60 & $3.41\text{\textbf e} \boldsymbol{ +06 }$ & $8.87\text{\textbf e} \boldsymbol{ +06 }$ &                                       --- & $3.72\text{\textbf e} \boldsymbol{ +01 }$ & $6.76\text{\textbf e} \boldsymbol{ +01 }$ &                                       ---\\ \hline
           &  80 & $1.95\text{\textbf e} \boldsymbol{ +06 }$ & $1.73\text{\textbf e} \boldsymbol{ +07 }$ &                                       --- & $4.58\text{\textbf e} \boldsymbol{ +01 }$ & $2.73\text{\textbf e} \boldsymbol{ +01 }$ &                                       ---\\ \hline
           & 100 & $2.90\text{\textbf e} \boldsymbol{ +06 }$ & $7.80\text{\textbf e} \boldsymbol{ +06 }$ &                                       --- & $5.22\text{\textbf e} \boldsymbol{ +01 }$ & $1.84\text{\textbf e} \boldsymbol{ +02 }$ &                                       ---\\ \hline
           & 250 & $2.08\text{\textbf e} \boldsymbol{ +06 }$ & $9.16\text{\textbf e} \boldsymbol{ +06 }$ &                                       --- & $1.06\text{\textbf e} \boldsymbol{ +02 }$ & $1.04\text{\textbf e} \boldsymbol{ +03 }$ &                                       ---\\ \hline
\end{tabular}

  \caption{Median values of 100 runs on all tested data sets and $k$. The first cost column for PROBI gives the cost on the summary, the second gives the cost on the full data set. The first PROBI column in the running time part gives the running time of PROBI, the second  indicates the additional time needed by \plloyd\ to compute the solution on the summary. The previous diagrams only show the PROBI time\label{table:all:median}.}
\end{centering}
\end{table}
\begin{table}%
\begin{centering}
  \begin{tabular}{|l||c|c|c|c|c|c|c|}
\hline
           &     & \multicolumn{3}{|c|}{Costs}     & \multicolumn{3}{|c|}{Running Time}\\ \hline
   Dataset &   k & \multicolumn{2}{|c|}{PROBI} & P-LLOYD++ & \multicolumn{2}{|c|}{PROBI} & P-LLOYD++\\ \hline \hline
  BigCross &  15 & $0.0155$ & $0.0312$ &       --- & $0.0092$ & $0.1509$ &       ---\\ \hline
           &  20 & $0.0095$ & $0.0067$ &       --- & $0.0020$ & $0.1162$ &       ---\\ \hline
           &  25 & $0.0060$ & $0.0037$ &       --- & $0.0032$ & $0.1514$ &       ---\\ \hline
           &  30 & $0.0101$ & $0.0037$ &       --- & $0.0023$ & $0.0545$ &       ---\\ \hline
           &  50 & $0.0110$ & $0.0021$ &       --- & $0.0031$ & $0.0204$ &       ---\\ \hline
           & 100 & $0.0074$ & $0.0017$ &       --- & $0.0016$ & $0.0082$ &       ---\\ \hline
           & 250 & $0.0058$ & $0.0016$ &       --- & $0.0153$ & $0.0128$ &       ---\\ \hline \hline
Caltech128 &  10 & $0.0038$ & $0.0009$ &       --- & $0.0056$ & $0.1411$ &       ---\\ \hline
           &  20 & $0.0036$ & $0.0006$ &       --- & $0.0037$ & $0.1210$ &       ---\\ \hline
           &  30 & $0.0019$ & $0.0003$ &       --- & $0.0051$ & $0.0598$ &       ---\\ \hline
           &  40 & $0.0012$ & $0.0004$ &       --- & $0.0023$ & $0.1311$ &       ---\\ \hline
           &  50 & $0.0016$ & $0.0002$ &       --- & $0.0025$ & $0.1001$ &       ---\\ \hline
           & 100 & $0.0017$ & $0.0003$ &       --- & $0.0015$ & $0.0993$ &       ---\\ \hline
           & 150 & $0.0009$ & $0.0001$ &       --- & $0.0016$ & $0.0674$ &       ---\\ \hline \hline
    Census &  10 & $0.0049$ & $0.0002$ &  $0.0000$ & $0.0033$ & $0.2906$ &  $0.0170$\\ \hline
           &  20 & $0.0032$ & $0.0001$ &  $0.0002$ & $0.0036$ & $0.2044$ &  $0.0152$\\ \hline
           &  30 & $0.0031$ & $0.0001$ &  $0.0003$ & $0.0038$ & $0.1562$ &  $0.0131$\\ \hline
           &  40 & $0.0043$ & $0.0004$ &           & $0.0018$ & $0.0802$ &          \\ \hline
           &  50 & $0.0022$ & $0.0002$ &  $0.0003$ & $0.0021$ & $0.0093$ &  $0.0169$\\ \hline
           &  75 & $0.0026$ & $0.0002$ &           & $0.0320$ & $0.0244$ &          \\ \hline
           & 100 & $0.0014$ & $0.0002$ &           & $0.0036$ & $0.0072$ &          \\ \hline \hline
 Covertype &  10 & $0.0224$ & $0.0044$ &  $0.0000$ & $0.0059$ & $0.1901$ &  $0.0655$\\ \hline
           &  20 & $0.0240$ & $0.0096$ &  $0.0008$ & $0.0079$ & $0.1430$ &  $0.0650$\\ \hline
           &  30 & $0.0126$ & $0.0071$ &  $0.0014$ & $0.0051$ & $0.0929$ &  $0.0497$\\ \hline \hline
     Tower &  20 & $0.1050$ & $0.1246$ &       --- & $0.0471$ & $0.1269$ &       ---\\ \hline
           &  40 & $0.0579$ & $0.0175$ &       --- & $0.0043$ & $0.0151$ &       ---\\ \hline
           &  60 & $0.0712$ & $0.0242$ &       --- & $0.0050$ & $0.0236$ &       ---\\ \hline
           &  80 & $0.0598$ & $0.1194$ &       --- & $0.0045$ & $0.0342$ &       ---\\ \hline
           & 100 & $0.0988$ & $0.0890$ &       --- & $0.0074$ & $0.0119$ &       ---\\ \hline
           & 250 & $0.0218$ & $0.0009$ &       --- & $0.0765$ & $0.0181$ &       ---\\ \hline
\end{tabular}

  \caption{Variance coefficients of 100 runs on all tested data sets and $k$. The first cost column for PROBI gives the cost on the summary, the second gives the cost on the full data set. The first PROBI column in the running time part gives the running time of PROBI, the second  indicates the additional time needed by \plloyd\ to compute the solution on the summary. The previous diagrams only show the PROBI time\label{table:all:var}.}
\end{centering}
\end{table}




\end{document}